\documentclass{article}
\usepackage{amsmath} 
\usepackage{amsfonts}
\usepackage{graphicx}
\usepackage{breqn} 
\usepackage{tikz}
\usetikzlibrary{decorations.markings}
\usetikzlibrary{calc}
\usepackage[all,cmtip]{xy}
\usepackage{tikz-cd}
\usepackage{enumitem}
\usepackage{cite}

\usepackage[margin=1in]{geometry}
\usepackage{hyperref}

\title{\Huge The n-point Exceptional  Universe}
\author{S. Farnsworth$^{a,b}$ \\ 
 	\small $^a$Department of Mathematics, University of Regensburg,\\ 
 	\small$^b$Max Planck Institute for Gravitational Physics (Albert Einstein Institute), Germany.
 }

\begin{document}

\maketitle

\begin{center}
	\textit{In memory of I.T. Todorov (1933-2025).}
\end{center} 

\vspace{.5cm}

	\begin{abstract}
		We solve an open problem in spectral geometry: the construction of finite-dimensional, discrete geometries coordinatized by non-simple, exceptional Jordan algebras.
 The approach taken is readily generalisable to broad classes of nonassociative  geometries, opening the door to the spectral geometric desciption of gauge  theories with exceptional symmetries.
We showcase a proof-of-principle 2-point geometry corresponding to the internal space of an $F_4\times F_4$ gauge theory with scalar content restricted by novel conditions arising  from the associative properties of the coordinate algebra. We then  formally establish a  setting for generalising to n-point exceptional Jordan geometries with distinct points coupled  via an action on 1-forms constructed as split Jordan bimodules.

	\end{abstract}
 
 	\section{Introduction: Why is  non-associative spectral  geometry  physically relevant, and what makes it technically  difficult?}
 
Spectral geometry, whether associative~\cite{Connes:1994kx,Connes:reconstruction,Connes:2008kx} or nonassociative~\cite{Farnsworth_2020,Boyle:2020,Boyle:2014wba,Wulkenhaar_1997,Hassanzadeh_2015,Akrami_2004,boylef4,carotenuto2019,DV:2016,DV:2019,Farnsworth:2013nza,Farnsworth:2014vva,Farnsworth:2020ozj,ShaneThesis,Besnard_2022}, is interesting for physics because it is well suited for describing finite-dimensional and discrete geometries. Such  geometries are useful for modeling the internal spaces of gauge theories~\cite{Chamseddine:2007oz} without encountering stability issues or a proliferation of unobserved dilaton fields~\cite{KKtheory}. They also provide novel possibilities for modelling the discretization of spacetime~\cite{marco1}. In this section we provide a conceptual outline of  what spectral geometry is, we explain why the nonassociative case is particularly relevant for modelling physics, and why a complete generalisation  to nonassociative spectral geometry has until recently been difficult to achieve.
In Section \ref{example_geometry}, we show that it is indeed possible to build non-trivial nonassociative geometries by explicitly constructing a  2-point geometry based on a non-simple exceptional Jordan algebra, which provides a geometric description of the internal space of an $F_4\times F_4$ gauge theory. In Section \eqref{jordmods} we introduce split Jordan bimodules together with bimodule homomorphisms, which will be  necessary for constructing  more general discrete geometries over exceptional Jordan algebras. In Section \eqref{sec_forms} we explain the construction of  Dirac operators and differential 1-forms in finite dimensional, discrete, exceptional Jordan, nonassociative, spectral geometries, before closing with a discussion in Section~\ref{sec_discussion}.

\subsection{What is spectral geometry?}

Spectral geometry~\cite{Connes:2008kx} generalises Riemannian geometry by encoding topological and metric data in algebraic structures.
A fundamental idea behind this approach is the deep correspondence between algebras and topological spaces.
One illustrative  example of this sort of correspondence appears in quantum mechanics: the space of pure states of a 2-state quantum system forms a sphere, known as the Bloch sphere~\cite{AlfsenShultz1998,Farnsworth_2020}. Mathematically, these pure states correspond to maps $\rho_\psi(a) = \langle \psi|a|\psi\rangle$, which take operators $a\in End(H)$ on the Hilbert space $H = \mathbb{C}^2$ and return complex numbers. The operator algebra of complex $2\times 2$ matrices, and its representation on the Hilbert space $H$ thus provides a tangible and explicit example of  a correspondence between an algebra on the one hand and a topological space on the other (i.e. the Bloch sphere).
  Spectral geometry~\cite{Connes:1994kx,Connes:reconstruction,Connes:2008kx}  generalises this idea, allowing a much broader class of topological spaces, including smooth manifolds, noncommutative spaces, and discrete spaces, to be described in terms of algebras represented on Hilbert spaces.

 The second key idea of spectral geometry is the correspondence between first-order derivative operators and metric data. Typically, the distance $d(x,y)$ between  two  points `$x$' and `$y$' on a Riemannian manifold is defined by the shortest path between them.
 However, there is an alternative way to define distance that doesn't rely on geodesics.
  Instead, one can search for a function $f$ such that $|f(x) -f(y)| = d(x,y)$, where the difference between the values of the function evaluated  at the two points equals the distance. Such a function always  exists because any function with a finite difference can be re-scaled to reproduce the geodesic distance. 
  The challenge is identifying an appropriate function for this purpose. To achieve this, we restrict attention to smooth functions whose gradients everywhere on the manifold are bounded by 1, and then search for one that maximizes the difference between the two points. Remarkably, this procedure recovers the standard notion of distance in Riemannian geometry. The key addition in spectral geometry is a `Dirac' operator $D$, which encodes the gradient, allowing this definition to be extended~\cite{Connes:Distance} to noncommutative, finite-dimensional, or discrete spaces, where the smooth manifold structure is absent but a corresponding `coordinate' algebra $A$ still provides the necessary topological information~\cite{Martinettidistance,Iochum:Krajewski:Martinetti}.

 If one absorbs the above foundational  `topological' and `metric' ideas, then one understands, at least conceptually, the basic reason why spectral geometries are classified by a `spectral triple' of data~\cite{Connes:1994kx,Connes:reconstruction,Connes:2008kx}:
  \begin{align}
 	T = (A,H,D)
 \end{align}
 where $A$ is an algebra that holds topological data, $D$ is a `Dirac' operator that contains metric data, and $H$ provides a concrete representation space for constructing the geometry. This packaging of geometric data is particularly valuable for modeling physical systems geometrically, as it offers a flexible framework for constructing a geometry that matches a physical system, rather than attempting to fit the system into the more restrictive framework of Riemannian geometry. 
 
  A prime example illustrating the advantages of spectral triples is the comparison between Kaluza-Klein theory~\cite{KKtheory} and Connes' noncommutative standard model~\cite{Chamseddine:2007oz}.  Kaluza-Klein theory attempts to model the internal structure of electromagnetism by compactifying a Riemannian space. However, the dynamics derived from the Einstein-Hilbert action lead to instabilities within the compactified internal space.
   The $5$-dimensional metric also gives rise to an unwanted dilaton field, in addition to the desired $4$-dimensional metric and a $U(1)$ gauge field. In contrast, while challenges remain with the noncommutative standard model, it does not suffer from these specific maladies.

 \subsection{Modelling physics with nonassociative geometry}
 
 \label{sec_modelling} 
 
Spectral geometry~\cite{Connes:1994kx,Connes:reconstruction,Connes:2008kx} traditionally assumes that the coordinate algebra $A$ of a spectral triple is associative. However, several key physical considerations motivate a more general framework that allows for nonassociative coordinate algebras.  These are:
\begin{enumerate}[wide, labelwidth=!, labelindent=0pt]
	\item \textbf{Physical Incompatibility with Associative Representations:} The fundamental input data of a spectral geometry consists of a coordinate algebra $A$ represented on a Hilbert space $H$.  In some cases, the natural Hilbert space associated with a physical system is not closed under the action of associative algebras. This issue arises most notably when the Hilbert space is constrained by anti-linear operators~\cite{Farnsworth_2020,Boyle:2020}.
	For instance, in quantum systems one may encounter Hilbert spaces of Hermitian $n\times n$ matrices, which are closed under the nonassociative Jordan product rather than the standard associative matrix product~\cite{townsandreview}. Similarly, Majorana spinors are subject to a constraint of the form $JH=H$, where $J$ is the anti-linear charge conjugation operator. Since $J$ does not generally commute with the associative action of complex matrix algebras on $H$, Majorana spinors will, in general, fail to remain closed under this associative action~\cite{Farnsworth_2020,Boyle:2020}.

\item \textbf{Chirality, and   Scalar Field Restrictions:} In an associative algebra $A$, the commutator with any element $a\in A$ defines a derivation:
\begin{align}
[a,bc] &= [a,b]c+b[a,c], & b,c&\in A\label{assocdir}
\end{align}
 This property has important consequences for associative spectral geometry.   When constructing a finite-dimensional spectral triple $T= (A,H,D)$, one typically requires that the Dirac operator $D$ acts as a derivation on the representation of the coordinate algebra~\cite{Dungen,landi1997}.  In the associative case, this condition is trivial, as any operator acts as a derivation under the commutator. Consequently, there is no natural mechanism restricting the form of $D$ arising from associativity.
 Because the scalar fields of a gauge theory are parameterized by the Dirac operator,  this  raises the question of why 
  we do not observe a proliferation of unwanted scalar fields, such as those absent in the Standard Model~\cite{Boyle:2014wba}. In the nonassociative setting, the situation is reversed: it becomes difficult to find an operator $D$ that acts as a derivation on the representation of the coordinate algebra. This aligns more naturally with the observed scarcity of scalar fields seen in reality. Additionally, the more restrictive nature of derivations in nonassociative algebras can lead to algebras that are inherently chiral with respect to their internal symmetries~\cite{toro}, a feature beneficial for modeling chiral gauge theories~\cite{Fredy} and one that does not seem to arise naturally in the associative setting.

	\item \textbf{Exceptional Symmetries:} The gauge symmetries of a spectral geometry correspond to the symmetries of the representation of the  coordinate algebra $A$ on the Hilbert space $H$~\cite{Dungen,Farnsworth_2015}. 	Notably, all algebras that exhibit exceptional symmetries are nonassociative. The symmetry group of the octonions is $G_2$ for example~\cite{baez}, while the exceptional jordan algebra has symmetry group $F_4$~\cite{boylef4}, and $E_6$ is the symmetry group of  Brown algebras~\cite{gari1}.  
	For this reason, nonassociative coordinate algebras are essential for modelling  physical systems with exceptional symmetry within the spectral geometry framework.

\end{enumerate}

In summary, nonassociative spectral geometry is not just an abstract generalisation but a framework that aligns more closely with key features of observed physics. It provides a natural setting for Hilbert spaces constrained by anti-linear operators, offers  built-in mechanisms for restricting scalar fields and modelling chiral symmetries, and is essential for describing exceptional symmetries that arise in certain models of high-energy physics~\cite{boylef4,todorov,tod2,Furey_2018}.

\subsection{The technical challenge of nonassociative geometry}

\label{sec_challenge}

Nonassociative geometries are technically  more challenging to implement than their associative counterparts. This difficulty is partly artificial, as significantly more research has been devoted to developing the associative framework (although see e.g. \cite{Farnsworth_2020,Boyle:2020,Boyle:2014wba,Wulkenhaar_1997,Hassanzadeh_2015,Barnes:2016cjm,Akrami_2004,boylef4,carotenuto2019,DV:2016,DV:2019,Farnsworth:2013nza,Farnsworth:2014vva,Farnsworth:2020ozj,ShaneThesis,Besnard_2022,Freudenthal} for previous work). However, there are also genuine technical obstacles unique to the nonassociative setting that must be overcome to construct viable nonassociative geometries.  This section outlines the central challenge, which the remainder of the paper will address.

To model a physical system, one needs both a parameterization of the system and a description of its dynamics. If this is to be achieved through an action principle, then defining a notion of differentiation, and of differential forms is essential.
The main challenge in building nonassociative spectral geometries lies in the technical difficulty of constructing Dirac operators and representations of differential 1-forms. Consider a nonassociative spectral triple $T= (A,H,D)$, where for now the exact properties of $A$, $H$, and $D$ are not specified, other than assuming they are as similar as possible to the associative setup, with the exception of $A$ being nonassociative. Assuming that one knows how to construct a nonassociative representation $\pi:A\rightarrow End(H)$, then one would like to use the Dirac operator to define a representation of `exact' 1-forms just as one does in the associative setting:
\begin{align}
	\pi(d[a]) := [D,\pi(a)]\in B(H),
\end{align}
for $a\in A$.  Further, as in the associative setting one would also  like to use the representation of the coordinate algebra, and of exact forms, to define a representation of more general 1-forms:
\begin{align}
	\pi(a d[b]) := \pi(a)\star \pi(d[b])= \pi(a)\star [D, \pi(b)]
\end{align}
where  $a,b\in A$. The difficulty arises in understanding what  is meant  by the product $\star:End(H)\times End(H)\rightarrow End(H)$ in the above expression. In the associative setting one can assume that the product $\star$  is given by the operator product, in which  case the Dirac operator acts as a derivation $[D,\pi(\cdot)]:A\rightarrow B(H)$ as required:
\begin{align}
	[D,\pi(ab)] = [D,\pi(a)]\star\pi(b)+ \pi(a)\star  [D,\pi(b)].\label{leib}
\end{align}
The above Leibniz rule holds in the associative setting when $\star$ is taken to be the operator product, not only for Riemannian geometries where $D$ is the curved space Dirac operator $\nabla^S$, but also for finite dimensional geometries where $D$ is essentially just a matrix. In the latter case the equality holds because  commutators act as derivations on associative algebras, including finite dimensional algebras of operators.  In the nonassociative setting, however, it is generally not possible to select $\star$ as the operator product, and commutators will in general no longer act as derivations~\cite{Schafer}.
This raises the fundamental question: under what conditions, if any, can one find a pair $(D,\star)$ for a given set of nonassociative order zero data $(A,H)$, such that the Leibniz rule in Eq.~\eqref{leib} holds? This challenge, which has no analogue in the associative setting, represents the key technical obstruction in nonassociative geometry. It makes nonassociative geometry more difficult, but also in many ways far more interesting than associative spectral geometry,  as it introduces novel restrictions on particle content when modeling gauge theories.

In Section \ref{example_geometry} we will construct an example  geometry based on the a non-simple exceptional Jordan algebra,  demonstrating that it is possible to construct discrete, nonassociative, finite-dimensional geometries, which have no analogue in the associative setting.  This example will also highlight the kinds of restrictions nonassociativity can impose on scalar content when modeling gauge theories. Subsequent sections will then provide a more rigorous treatment of discrete geometries based on exceptional Jordan algebras.

\section{The two-point exceptional Jordan Geometry}
\label{example_geometry}

The geometries most relevant for modeling the internal spaces of gauge theories involve finite-dimensional, non-simple algebras \( A = A_0 \oplus \dots \oplus A_n \)~\cite{Connes:2008kx}. The factors \( A_i \) correspond to distinct points in a discrete geometry, and are decoupled from one another at the level of the algebraic product and sum, which act point-wise:
\begin{align}
	(a_1,...,a_n)(b_1,...,b_n) &= (a_1b_1,...,a_nb_n)\\
		(a_1,...,a_n)+(b_1,...,b_n) &= (a_1+b_1,...,a_n+b_n)
\end{align}
$a=	(a_1,...,a_n),b=	(b_1,...,b_n)\in A$. It is only through the introduction of differential 1-forms (and other higher order data) that the factors of a non-simple coordinate algebra become coupled, as we illustrate schematically:
\begin{align}
a(\omega b) = 
\begin{pmatrix}
	a_1 & 0 & 0 & 0 \\
	0 & a_2 & 0 & 0 \\
	0 & 0 & \ddots & 0 \\
	0 & 0 & 0 & a_n
\end{pmatrix}
\left(
\begin{pmatrix}
	0 & \omega_{12} & \omega_{13} & \dots \\
	\omega_{21} & 0 & \omega_{23} & \dots \\
	\omega_{31} & \omega_{32} & 0 & \dots \\
	\vdots & \vdots & \vdots & \ddots
\end{pmatrix}
\begin{pmatrix}
	b_1 & 0 & 0 & 0 \\
	0 & b_2 & 0 & 0 \\
	0 & 0 & \ddots & 0 \\
	0 & 0 & 0 & b_n
\end{pmatrix}\right),\label{left-right}
\end{align}
where $\omega\in \Omega_D^1 A$ is a generic `1-form'. Such geometries, with finite dimensional, non-simple coordinate algebras coupled at higher order,
correspond to the internal spaces of gauge theories with multiple gauge sectors connected by Higgs fields~\cite{Dungen,Chamseddine:2007oz}.
For this reason it is these kinds of  spaces that are of primary interest when  extending the spectral geometry formalism to include nonassociative algebras, and it is therefore these spaces which we explore  now.

In previous work~\cite{Boyle:2020,Besnard_2022} we demonstrated how to construct discrete, finite dimensional, non-associative geometries for the particular case in which the coordinate algebra is a special Jordan algebra. However, since special Jordan algebras can always be embedded in associative algebras, it is important to construct proof of principle examples that are intrinsically nonassociative.  Here we present such an example geometry $T = (A,H,D)$, with coordinate algebra $A = J_3(\mathbb{O})\oplus J_3(\mathbb{O})$, where each factor $J_3(\mathbb{O})$ is a copy of the exceptional Jordan algebra. As with all algebras explored in this paper, this is a nonassociative, finite dimensional, real, unital algebra,  (see Appendix Section \ref{sec_jorda} for a brief  overview of the exceptional Jordan algebra).
         Before constructing a  geometry coordinatized by $A$, however, it should be pointed out that the geometry coordinatized by a single copy of the exceptional Jordan algebra $J_3(\mathbb{O})$ is well understood. The space of pure states is the 
         octonionic projective plane, denoted $\mathbb{OP}^2$, which is a 16 dimensional real manifold. We provide a brief overview of this space in Appendix \ref{sec_state}, and more in depth reviews can be  found elsewhere~\cite{baez}. Furthermore, a beautiful `derivation' based calculus for $J_3(\mathbb{O})$ has been developed in~\cite{carotenuto2019,DV:2016}.  
Confident that the geometry based on a single copy of the exceptional Jordan algebra makes sense, we focus on the geometry of two discrete points with a Dirac operator encoding the metric data ``between'' them.

All representations of the exceptional Jordan algebra are free~\cite{jacob1968}, and as such we take the Hilbert space to be $H = J_3(\mathbb{O})\otimes\mathbb{R}^2$, with a natural action of the algebra  $A = J_3(\mathbb{O})\oplus J_3(\mathbb{O})$ given by:
\begin{align}
	\pi(a)h = \begin{pmatrix}
		S_{a_1}&0\\
		0&S_{a_2}
	\end{pmatrix}\begin{pmatrix}
	h_1\\
	h_2
\end{pmatrix}\label{Exceptional_Rep} 
\end{align} 
where $a=(a_1,a_2)\in A$, $h=(h_1,h_2)\in H$, and where $S_ab = ab$, $a,b\in J_3(\mathbb{O})$ denotes the action of the exceptional Jordan algebra on itself. The representation `$\pi$' is symmetric with respect to the inner product on $H$, which we review  in Appendix Sections \ref{sec_state} and \ref{sec_2pointdistance}. We  also provide a more formal description of Jordan modules in section \ref{sec_mod_gen}. 
For now, note that even though the exceptional Jordan algebra is nonassociative,  the objects $S_a$ are standard operators, which can be expressed as $S_a = (S_a)^{ij}e_i\otimes e_j^\ast$ where the $e_i$ form a $27$-dimensional basis for $J_3(\mathbb{O})$, and `$\ast$'  denotes the dual,  with $e_i^\ast S_a = (ae_i)^\ast$.
It is important to remind the reader that in general $S_{ab}e_i = (ab)e_i\neq a(be_i)=S_aS_be_i $. The only algebra element that associates with all other elements of $J_3(\mathbb{O})$ is the identity element $e_0$, such that $S_{ab}e_0 = (ab)e_0 = a(be_0) = S_aS_be_0$, $\forall a,b\in J_3(\mathbb{O})$.

  Having introduced the data $(A,H)$, the next step is to complete our spectral triple by constructing  a Dirac operator $D$ that is compatible with the representation of the coordinate algebra and its symmetries.
    What we would like is a Dirac operator that `connects' the two, otherwise independent, points in the geometry (i.e. the two factors in the algebra). The most general Hermitian operator on $H$ that does this takes the form:
\begin{align}
	D = M^{ij}\begin{pmatrix}
		0 & e_i\otimes e_j^\ast\\
		 e_j\otimes e_i^\ast   &0
	\end{pmatrix} 
\end{align}
where the $M^{ij}$ are real coefficients.  Of course, we don't want the most general operator, but rather a derivation satisfying  Eq.~\eqref{leib}. Making use of our representation, this equation becomes:
	\setlength{\arraycolsep}{0pt}
\begin{align}
	LHS&= [D,\pi(ab)]\nonumber\\
	&=M^{ij}\begin{pmatrix}
		0 & e_i\otimes ((a_2b_2)e_j)^\ast - (a_1b_1)e_i\otimes e_j^\ast \\
e_j\otimes ((a_1b_1)e_i)^\ast  -(a_2b_2)e_j\otimes e_i^\ast &0
	\end{pmatrix}\nonumber\\
RHS&=[D,\pi(a)]\star\pi(b)+ \pi(a)\star  [D,\pi(b)]\nonumber\\
	&=M^{ij}\begin{pmatrix}
	0 & e_i\otimes (a_2e_j)^\ast - a_1e_i\otimes e_j^\ast \\
	e_j\otimes (a_1e_i)^\ast  -a_2e_j\otimes e_j^\ast &0
\end{pmatrix}\star\pi(b)\nonumber\\
&+M^{ij}\pi(a)\star\begin{pmatrix}
	0 & e_i\otimes (b_2e_j)^\ast - b_1e_i\otimes e_j^\ast \\
	e_j\otimes (b_1e_i)^\ast  -b_2e_j\otimes e_i^\ast &0
\end{pmatrix}\label{comparison}
\end{align}
for $a,b\in A$, and where $D$ and $\pi(a)$ are standard operators in $End(H)$. Similarly elements of the form $[D,\pi(a)]\star\pi(b)$ are operators in $End(H)$, but we have not yet assumed anything about the product $\star:End(H)\times End(H)\rightarrow End(H)$ between the representation of zero forms and 1-forms. In particular we have not assumed associativity, and as such one has in general:
\begin{align}
	\pi(ad[b])h:=(\pi(a)\star [D,\pi(b)])h\neq \pi(a)[D,\pi(b)]h.
\end{align}
For simplicity, however, let us now make a rather counter-intuitive assumption: let us \textit{assume} that  $\star$ is given  by the associative operator product. The justification for this choice will be discussed in Section \ref{sec_mod_exce}, where we introduce Jordan bimodules in a formal manner. For now, notice that if this assumption is taken then  the  RHS of 
 Eq.~\eqref{comparison} becomes: 
\begin{align}
	RHS&=M^{ij}\begin{pmatrix}
		0 & e_i\otimes (b_2(a_2e_j))^\ast - a_1(b_1e_i)\otimes e_j^\ast \\
		e_j\otimes (b_1(a_1e_i))^\ast -a_2(b_2e_j)\otimes e_j^\ast &0
	\end{pmatrix}
\end{align}
Remembering that the exceptional Jordan algebra is commutative, we see that the RHS and LHS are almost identical, except for bracket ordering in the off-diagonal components. Unlike in the associative setting, assuming $\star$ to be the associative operator product \textit{does} place restrictions on the form of $D$. In particular, it forces all coefficients $M^{ij}$ to be zero, except for  $M^{00}$. This is because the identity element $e_0$ of the exceptional Jordan algebra is the only element that associates with every other basis element in the algebra allowing us to satisfy $LHS=RHS$ with $M^{00}\neq 0$. The assumption that $\star$ is given by the operator product therefore forces us to take:
\begin{align}
	D = \kappa\begin{pmatrix}
		0 & e_0\otimes e_0^\ast\\
		e_0\otimes e_0^\ast  &0
	\end{pmatrix} \label{Dirac}
\end{align}
for some $\kappa\in \mathbb{R}$. 

Having constructed the data of our spectral triple $T=(A,H,D)$, the next step is to ensure that the Dirac operator is compatible with the symmetries of the representation. The symmetries of the representation are generated by `inner' derivation elements of the form~\cite{Boyle:2020,Schafer,Besnard_2022}:
\begin{align}
	\delta_{a,b}&=[\pi(a),\pi(b)]=\begin{pmatrix}
		[S_{a_1},S_{b_1}]&0\\
		0&[S_{a_2},S_{b_2}]
	\end{pmatrix} \label{excep_der}
\end{align}
$a,b\in A$. These derivation elements form a representation of the $104$ dimensional Lie algebra $L =f_4\oplus f_4$, which  generates the symmetry group $G = F_4\times F_4$ of the geometry~\cite{Ramond}.  This symmetry group acts independently on the `points' of our 2-point, exceptional Jordan geometry. Because the identity element of the Exceptional Jordan algebra is invariant under the automorphism group $F_4$,  the Dirac operator defined in Eq.\eqref{Dirac} naturally commutes with the derivation elements:
\begin{align}
	[D,\delta_{a,b}] = 0
\end{align}
for all $a,b\in A$. As such the Dirac operator is compatible with the symmetries of the representation, but it doesn't fluctuate. This is similar to what  occurs for the $\sigma$ field in the noncommutative standard model~\cite{Chamseddine:2012fk}. Were we to construct a full gauge theory by taking the product of our internal space `$T$' with a Riemannian geometry, we would obtain an $F_4\times F_4$ gauge theory with Fermions in fundamental  and singlet representations, and an uncharged scalar field (resulting from the single free parameter `$\kappa$' in \eqref{Dirac}) coupling the `left' and `right' Fermion sectors. We won't  perform this calculation here, as it is fairly mechanical and similar calculations for nonassociative geometries based on the octonions and special Jordan algebras can be found elsewhere~\cite{Farnsworth:2013nza,Besnard_2022}.

Given the Dirac operator specified in \eqref{Dirac}, together with our  assumption for the product between zero and 1-forms, a general `1-form', $\omega\in \Omega_D^1 A$ will take the form:
\begin{align}
\omega	=\sum\pi(a_1)...\pi(a_{n-1})[D,\pi(a_n)],\label{general-1-form}
\end{align}
in which the sum is over the elements $a_i\in A$. Because the coordinate algebra $A$ is finite dimensional, any 1-form can be expressed such that the chain length `$n$' in each term  of the above sum remains finite. We remind the reader that in the above expression $\pi(a)\pi(b)\neq \pi(ab)$ in general. However one always has $[D,\pi(ab)] = [D,\pi(a)\pi(b)]$, which together with the Leibniz rule is what allows us to express a general 1-form as in \eqref{general-1-form} with the action of the coordinate algebra coming only from the left.

Finally, the  distance between the two points in this example geometry is well defined in terms of Connes distance function. In appendix section \ref{sec_2pointdistance}, we go through this calculation, and show that it is given by:
\begin{align}
	d(x,y) = \frac{1}{\kappa}
\end{align}
where $\kappa$ is the coefficient defined in \eqref{Dirac}, and the points `$x$' and `$y$' correspond to pure states acting on  the two independent factors of the algebra respectively. This is the  same answer one would expect to obtain for a 2-point geometry in the associative setting~\cite{Martinettidistance,Iochum:Krajewski:Martinetti}.

	\section{Representations of Jordan Algebras}
	\label{jordmods}

The example 2-point  geometry introduced in section \ref{example_geometry} was constructed using  two kinds of modules. First, the Hilbert space  $H$ acted as a module over the coordinate algebra $A$, and over the one forms $\Omega_D^1 A$. Secondly, the space of 1-forms $\Omega_D^1 A$ acted as a `bimodule' over the coordinate algebra $A$. More generally, modules appear throughout  the construction of spectral geometries. Because the representation theory for nonassociative algebras~\cite{Eilenberg} is less well developed than that of associative algebras, it is important to establish clear definitions of what one means by  representations if one wishes to make progress in the nonassociative setting. This is the focus of the current section.  While there is a standard definition for bimodules over Jordan algebras~\cite{jacob1968,carotenuto2019}, the space of 1-forms in the example geometry given in section 
\ref{example_geometry}  fits a slightly more general definition than what is usually considered.  In subsection \ref{sec_mod_gen} below we clarify what is meant by a Jordan bimodule. Subsection \ref{sec_mod_exce} then focuses on the specific case of bimodules over exceptional Jordan algebras, while subsection \ref{sec_mod_maps} introduces module maps on Jordan bimodules, which will be useful when constructing differential graded algebras of forms.

	\subsection{Jordan Bimodules} 
	\label{sec_mod_gen}
	Let $A$ be a real, unital, Jordan algebra, and $\pi:A\rightarrow End(M)$ be a linear map from $A$ into the endomorphisms on  a  vector space $M$ (see \cite{Schafer,jordanneumannwigner} for an overview of Jordan algebras). Then $(M,\pi)$ is said to be a Jordan module over $A$ if $\pi$ satisfies~\cite{jacob1968,Besnard_2022}:
	\begin{align}
		[\pi(a),\pi(a^2)]&=0,\label{mult1}\\
		2\pi(a)\pi(b)\pi(a)+\pi(a^2\circ b)&=2\pi(a\circ b)\pi(a)+\pi(a^2)\pi(b)\label{jordact}
	\end{align}
$a,b\in A$, and where $\circ:A\times A\rightarrow A$ corresponds to the Jordan product on $A$.
 	 	 A Jordan bimodule is a vector space $M$ that is equipped with two linear maps $\pi_L$ and $\pi_R$, which each obey the above properties, and which are `compatible' in a sense that we will now explain. The usual approach to compatibility is to define a Jordan bimodule such that the left and right actions are identified: $\pi_L=\pi_R$~\cite{DV:2016,carotenuto2019}.
	When this identification is made,  the definition of a   Jordan bimodule  $M$ over an algebra $A$ is equivalent to defining a new algebra $B= A\oplus M$ equipped with the following product:
	\begin{align}
		(a+m)(a'+m')&=aa'+ (am'+ma')\nonumber\\
		&=aa'+\pi_L(a)m'+\pi_R(a')m\label{bimodule}
	\end{align}
	$a,a'\in A$, $m,m'\in M$, and where   $B$  satisfies all of the properties of a Jordan algebra\cite{Eilenberg,Kashuba}.

	In this paper we make use of a slightly  more general setup for Jordan bimodules in which $\pi_L\neq\pi_R$, but where  the  following compatibility condition is satisfied:
	\begin{align}
		[\pi_L(A),\pi_R(A)]=0.\label{splitcomm}
	\end{align}
	In this case the algebra $B=A\oplus M$ equipped with the product given in Eq.~\eqref{bimodule} no longer satisfies the  properties of a Jordan algebra in general because the product is no longer symmetric when acting on elements of $M$. Throughout this paper we will stick with the standard terminology, referring to any Jordan bimodule in which the left and right actions are identified simply as a Jordan bimodule. \textit{Any Jordan  bimodule for which $\pi_L\neq \pi_R$ and where Eq.~\eqref{splitcomm} is satisfied will be called a `split' Jordan bimodule.} 
	Wherever confusion might arise we will use the terminology `non-split' to remind the reader that we are referring to the canonical  definition of a Jordan bimodule in which $\pi_L=\pi_R$. 
	
	Our goal in introducing split Jordan bimodules is the construction of differential graded algebras $\Omega_d A = \oplus_i \Omega^i_d A$  defined at order zero by  non-simple exceptional Jordan algebras $\Omega_d^0A :=A=\oplus_1^n J_3(\mathbb{O})$. Note, however, that the canonical definition of a `non-split' Jordan bimodule doesn't exclude the construction of Jordan differential graded algebras. A beautiful derivation based calculus was defined over a single copy of the exceptional Jordan algebra in~\cite{DV:2016,carotenuto2019}, based on the standard Jordan bimodule definition. In the current  paper, however, we are not interested in the geometry of a single exceptional point, but rather in establishing metric data `between' the points of a discrete exceptional geometry.  In~\cite{Besnard_2022} we  introduced a construction of 1-forms  capable of dealing with discrete, special Jordan geometries 
 using the standard `un-split' definition of Jordan bimodules. This raises the obvious question: why not use our already established construction when dealing with non-simple exceptional Jordan algebras? What is it that goes wrong?

  For special Jordan algebras one is able to take advantage of the fact that Jordan bimodules can be constructed from what are known as `associative' representations. Specifically, for special Jordan algebras one is able to construct `associative' representations $\pi:A\rightarrow End(M)$, satisfying:
\begin{align}
	\pi(a\circ b) = \frac{1}{2}(\pi(a)\pi(b)+\pi(b)\pi(a))\label{assocrep}
\end{align}
for $a,b\in A$, where $\circ$ is the Jordan product on $A$,  and where juxtaposition on the right hand side   indicates the usual operator product. Given any two associative representations $\pi_L$ and $\pi_R$ of a special Jordan algebra $A$ satisfying $[\pi_L(A), \pi_R(A)] = 0$, one can always construct a new `symmetric' representation:
\begin{align}
\pi_S = \frac{1}{2}(\pi_L + \pi_R)\label{symmetricact}
\end{align}
which satisfies the Jordan module properties given in Eqs. \eqref{mult1} and \eqref{jordact}~\cite{Besnard_2022}. When dealing with special Jordan algebras, this fact allows one to construct a space of one-forms $\Omega_D^1A$   as a standard un-split Jordan bimodule, with a symmetric   action composed of  associative actions~\cite{Besnard_2022}. The two `associative' actions  `$\pi_L$' and `$\pi_R$' of $A$ on $\Omega_D^1 A$
 function like the two ends of a bridge, linking the points in a discrete geometry. 

 To see  how this works in practice,  consider a spectral triple $T=(A,H,D)$ with special Jordan algebra $A$, and with  an `associative' representation $\pi:A\rightarrow End(H)$. 
  To define a consistent action of $A\oplus \Omega_D^1 A$ on $H$, we extend the associative representation rule in Eq. \eqref{assocrep} to include differential 1-forms, leading to the following product rules~\cite{Besnard_2022}:
\begin{align}
	\pi(a)\star\pi(b)&= \pi(a\circ b) = \frac{1}{2}(\pi(a) \pi(b)+ \pi(b)\pi(a) )\label{formsproductspecial0}\\
\pi(a)\star\pi(d[b])&= \pi(a\circ d[b]) = \frac{1}{2}(\pi(a) \pi(d[b])+ \pi(d[b])\pi(a) )\label{formsproductspecial}
\end{align}
$a,b\in A$. This product is automatically consistent with the  Leibniz rule in Eq.~\eqref{leib}. 

While the action of $A$ on $H$ is associative,  the action of $A$ on itself and  on $\Omega_D^1A$, as given by $\star$, is  symmetric and satisfies Eqs.~\eqref{mult1} and \eqref{jordact}, making it a multiplicative representation. That is, the space of 1-forms acts as a non-split Jordan bimodule over $A$.  One sees directly from Eqs.~\eqref{formsproductspecial0} and~\eqref{formsproductspecial} that the `left' and right associative actions from which the multiplicative action $\star$ is composed, are nothing other than the left and right operator products.  
This setup allows one to construct finite dimensional and discrete special Jordan geometries, where one factor of a non-simple coordinate algebra $A=\oplus_i^n A_n$ might act on a given component of a 1-form from the left, while another factor acts from the right, as shown schematically in Eq.~\eqref{left-right}, thereby providing a space for otherwise disconnected sectors of a geometry to interact, just as in the associative setting. 

Unfortunately, while the above construction works for special Jordan algebras, for exceptional Jordan algebras, associative representations do not exist, 
making our original  construction~\cite{Besnard_2022} based on associative representations infeasible. We introduce  split Jordan bimodules because they provide a structure that recovers certain desirable features of multiplicative representations constructed from associative representations, allowing one to construct first order data that couples  the discrete points of an exceptional Jordan geometry. Namely, one is able to begin with a multiplicative representation of $A$ on $H$, which then  gives rise naturally to a split Jordan bimodule structure  on $\Omega_D^1 A$, in exactly the same way that one  arrives naturally at a non-split bimodule structure on $\Omega_D^1 A$ if one begins with an associative representation of $A$ on $H$. We formalise this idea in the sections that follow.

.




\subsection{Bimodules Over  Exceptional Jordan Algebras}
\label{sec_mod_exce}
For the exceptional Jordan algebra $A=J_3(\mathbb{O})$, any finite-dimensional non-split Jordan bimodule $M$ is a direct sum of bimodules isomorphic to $A$~\cite{jacob1968}. That is, one can always write $	M = J_3(\mathbb{O})\otimes V$, 
where $V$ is some finite-dimensional real vector space~\cite{jacob1968}, and the action of the algebra is given in the obvious way by $\pi(a)(h\otimes v) = ah\otimes v$. More generally, for non-simple exceptional Jordan algebras with $n$ factors:
\begin{align}
	A=\oplus^n_1 J_3(\mathbb{O}),\label{discretealg}
\end{align}
the corresponding Jordan bimodules take the form:
\begin{align}
M = J_3(\mathbb{O})\otimes \left(\oplus^n_{i=1}V^i\right), \label{initjord}
\end{align}
where each $V^i$ is a real vector space, and the  algebra acts via:
\begin{align}
\pi(a)(h\otimes v) =\sum^n_{i=1} a_ih\otimes P^i(v),\label{nonsplitaction}
\end{align}
for $a= (a_1,...,a_n)\in A$, and  where the $P^i$ are projection operators onto the subspaces $P^i(V) = V^i$. 

While  standard non-split bimodules correspond to the case in which the left and right actions of the algebra are identified, split Jordan bimodules establish separate left and right actions satisfying Eqs.~\eqref{mult1}, \eqref{jordact}, and \eqref{splitcomm}.  We introduce split Jordan bimodules of the form:
\begin{align}
M = J_3(\mathbb{O})\otimes (\oplus_{i,j=1}^nV^{ij})\otimes J_3(\mathbb{O})\label{newmoduleform}
\end{align}
where the $V^{ij}$ are  real sub-vector spaces. The coordinate algebra $A$ over which $M$ is defined, still has $n$ discrete factors as in Eq.~\eqref{discretealg}. The left and right actions of the algebra are defined  as:
\begin{align}
	\pi_L(a)(h\otimes v\otimes k) & = \sum_{ij}^n a_i h \otimes P^{ij}(v)\otimes k\label{mult11}\\
		\pi_R(a)(h\otimes v\otimes k) & = \sum_{ij}^n  h \otimes P^{ij}(v)\otimes a_jk\label{mult2}
\end{align}
where $a = (a_1,...,a_n)\in A$, and  $P^{ij}$ projects onto the subspace $P^{ij}(V) = V^{ij}$. 

Revisiting the example  geometry $T=(A,H,D)$ from section \ref{example_geometry},
 the coordinate algebra was given by $A = J_3(\mathbb{O})\oplus J_3(\mathbb{O})$,  while the Hilbert space $H = J_3(\mathbb{O})\otimes \mathbb{R}^2$ was a standard non-split Jordan bimodule of the form given in Eq.~\eqref{initjord}  
  with an algebra action given as in Eq.~\eqref{nonsplitaction}. On the other hand, the space of 1-forms $\Omega_D^1 A =  J_3(\mathbb{O})\otimes \mathbb{R}^2\otimes  J_3(\mathbb{O})$ was a split  Jordan bimodule of the form given in Eq.~\eqref{newmoduleform}, with $V^{21} = \mathbb{R}$, $V^{12} = \mathbb{R}$, and $V^{11} = V^{22} = 0$, with  a general element  $ \omega\in \Omega_D^1 A$ represented as:
\begin{align}
	\omega = \sum\begin{pmatrix}
		0 & \omega_1\otimes \omega_2^\ast\\
		\omega_3\otimes \omega_4^\ast  &0
	\end{pmatrix}.
\end{align}
$\omega_i\in J_3(\mathbb{O})$. The  left and right actions were assumed to be of the form:
\begin{align}
	\pi(a)\star\omega := \pi(a)\omega &=\sum\begin{pmatrix}
		0 & a_1\omega_1\otimes \omega_2^\ast\\
		a_2\omega_3\otimes \omega_4^\ast  &0
	\end{pmatrix}\label{action11}\\
	\omega \star\pi(a) := \omega\pi(a) &=\sum\begin{pmatrix}
	0 & \omega_1\otimes (a_2\omega_2)^\ast\\
	\omega_3\otimes (a_1\omega_4)^\ast  &0\label{action22}
\end{pmatrix}
\end{align}
for $a=(a_1,a_2)\in A$, which matches the form given in Eqs.~\eqref{mult11} and \eqref{mult2}. It now becomes apparent that  when we assumed the  product $\star$ to be the standard associative  operator product, the operators `$S_a$' defining the algebra representation on $H$, naturally extended to an action on 1-forms, inducing the standard Jordan action. In this light, our assumed action no longer appears so counterintuitive as is may first have seemed.

\subsection{Module Maps}
\label{sec_mod_maps}

Bimodule homomorphisms are used in associative spectral geometry when defining  representations of 1-forms. They play a similarly useful role in the  Jordan setting, and so we introduce them here.  Given two left modules $M$ and $N$ over a (possibly nonassociative) algebra $A$, a module homomorphism between $M$ and $N$ is a linear map $\phi:M\rightarrow N$, which satisfies the following rule:
\begin{align}
	\pi_N(a)\phi(m) = \phi(\pi_M(a)m),\label{modulemap}
\end{align}
for all $a\in A$, and $m\in M$. In the above expression $\pi_M:A\rightarrow End(M)$, and $\pi_N:A\rightarrow End(N)$ indicate the representations of $A$ on $M$ and $N$ respectively. In what follows, however, this notation will become too cumbersome, and so we will instead simply use juxtaposition to indicate the action of algebra elements on module elements. With this convention in place, Eq. \eqref{modulemap}  becomes:
\begin{align}
	a\phi(m) = \phi(am).\label{modulemap2}
\end{align}
The generalisation to bimodule maps is straightforward.
Given two  bimodules $M$ and $N$ over an algebra $A$, a bimodule homomorphism between $M$ and $N$ is a linear map $\phi:M\rightarrow N$, which satisfies the properties of a module homomorphism from both the left and right such that in addition to \eqref{modulemap2}, we also have:
\begin{align}
	\phi(m)a&:=\phi(ma).
\end{align}
$\forall a\in A$, $m\in M$.

Homomorphisms between finite dimensional, Jordan bimodules $M = A\otimes V$ and $N = A\otimes W$  over the exceptional Jordan algebra $A = J_3(\mathbb{O})$ are always of the form~\cite{carotenuto2019}:
\begin{align}
	\phi(a\otimes v) = a\otimes \Gamma w\in N
\end{align}
for $a\otimes v\in M$, $w\in W$, and where $\Gamma: V\rightarrow W$ is a linear map. To see this, notice first of all  that:
\begin{align}
	\phi(a(b\otimes v)) &= \phi(ab\otimes v)\nonumber\\
	&=(ab)\phi(e^0\otimes v_i)\label{mmm1}
\end{align}
$a\in A$, $b\otimes v\in M$, and where $e^0$ is the identity element in $J_3(\mathbb{O})$. For the second equality we have used the  module map property given in \eqref{modulemap2}. Making use of this identity again however, we also have
\begin{align}
	\phi(a(b\otimes v))&= a\phi(b\otimes v)\nonumber\\
	&= a(b \phi(e^0\otimes v))\label{mmm2}
\end{align}
Comparing Eqs.~\eqref{mmm1} and \eqref{mmm2} then tells us that $\phi(e^0\otimes v)\propto e^0\otimes v'$, for some $v'\in W$, because $e^0$ is the only element  that associates with all other elements of $J_3(\mathbb{O})$. But since we can always use the module map property \eqref{modulemap2} to write $\phi(a\otimes v) = a\phi(e^0\otimes v)$, we therefore have:
\begin{align}
	\phi(a\otimes v_i) &= a \otimes \Gamma_i^jw_j,
\end{align}
where the $\Gamma^i_j$ are real coefficients, and the $v_j$ and $w_j$ form a basis for $V$ and $W$ respectively. 

The generalisation to module homomorphisms between split Jordan bimodules in which the left and right actions are no longer identified follows naturally. Given two split Jordan bimodules 
 of the form $M = J_3(\mathbb{O})\otimes V\otimes J_3(\mathbb{O})$ and $N = J_3(\mathbb{O})\otimes W\otimes J_3(\mathbb{O})$ over  a finite dimensional, discrete exceptional Jordan algebra with $n$ discrete factors $A = \oplus^n_1 J_3(\mathbb{O})$, a bimodule homomorphism $\phi: M\rightarrow N$ will always take the form:
\begin{align}
\phi(a\otimes v\otimes b) = a\otimes \Gamma w \otimes b\label{homotran}
\end{align} 
Where $\Gamma: V\rightarrow W$ is a linear map. The proof is a little bit more involved than when dealing with Jordan bimodules over an exceptional Jordan algebra with a single factor, but the logic is the same. From the left one has:
\begin{align}
	a(b(\phi(e_0\otimes v\otimes c))) & = a(\phi(b(e_0\otimes v\otimes c)))\nonumber\\
	&=\sum_{ij}^n a(\phi(b_i\otimes P^{ij}(v)\otimes c))\nonumber\\
	&=\sum_{ij}^n \phi(a(b_i\otimes P^{ij}(v)\otimes c))\nonumber\\
		&=\sum_{kl}^n\sum_{ij}^n \phi(a_kb_i\otimes P^{kl}(P^{ij}(v))\otimes c)\nonumber\\
				&=\sum_{ij}^n \phi(a_ib_i\otimes P^{ij}(v)\otimes c)\label{LHScomp}
\end{align}
for $a= (a_1,...a_n),b=(b_1,...,b_n)\in A$, and 
where we have used the fact that projection operators satisfy $P^{kl}P^{ij} = \delta^{ki}\delta^{lj}P^{ij}$. On the other hand one has:
\begin{align}
	(ab)(\phi(e_0\otimes v\otimes c)) & = \phi((ab)(e_0\otimes v\otimes c))\nonumber\\
	&=\sum_{ij}^n \phi(a_ib_i\otimes P^{ij}(v)\otimes c)\label{LHScomp2}
\end{align}
Comparing \eqref{LHScomp} and \eqref{LHScomp2}, then yields:
\begin{align}
	(ab)(\phi(e_0\otimes v\otimes c))=a(b(\phi(e_0\otimes v\otimes c))), 
\end{align}
for all $a,b,c\in J_3(\mathbb{O})$, and $v\in V$. This implies that  $\phi (e_0\otimes v\otimes a)\propto (e_0\otimes v'\otimes a')\in N$. A similar argument from the right also yields $\phi (a\otimes v\otimes e_0)\propto (a'\otimes v'\otimes e_0)\in N$. We therefore have:
\begin{align}
	\phi (e_0\otimes v\otimes e_0) =  e_0 \otimes \Gamma w\otimes e_0
\end{align}
where  $\Gamma$ is a linear map from $V$ into  $W$. Next, we introduce the following notation: $a^{(k)} = (0,...,a,...,0)\in A$, where only the $k^{\text{th}}$ element of the algebra is non-zero. We then have:
\begin{align}
	\phi (a\otimes P^{ij}(v)\otimes b) &= \phi (a^{(i)}(e_0\otimes v\otimes e_0)b^{(j)})\nonumber\\
	&= a^{(i)} (\phi (e_0\otimes v\otimes e_0))b^{(j)}\nonumber\\
		&= a^{(i)} (e_0\otimes \Gamma w\otimes e_0)b^{(j)}\nonumber\\
				&= a\otimes P^{ij}(\Gamma w)\otimes b.\label{discretmodorpph}
\end{align}
This tells us that not only do bimodule homomorphisms between split Jordan bimodules over non-simple exceptional Jordan algebras take the form given in \eqref{homotran}, but further, the transformation $\Gamma:V\rightarrow W$ respects the split bimodule decomposition of the vector spaces $V$ and $W$.

\section{Differential 1-Forms}		

\label{sec_forms}

In associative  spectral geometry~\cite{Connes:1994kx,Connes:reconstruction,Connes:2008kx}, differential 1-forms are constructed by first building what is known as the space of  universal differential graded  1-forms $\Omega_d^1 A$ over a  coordinate algebra $A$~\cite{landi1997}. Connes'  1-forms $\Omega_D^1 A$ corresponding to a geometry $T=(A,H,D)$, are then constructed as a representation of the universal 1-forms using an appropriate bimodule homomorphism  $\pi:\Omega_d^1 A\rightarrow \Omega_D^1 A\subset End(H)$. In section \ref{universalforms} we review the key ideas behind the construction of Connes' 1-forms, and showcase certain features of associative bimodule maps, which are not often discussed, and which distinguish the associative and nonassociative settings from one another. In section \ref{explicdga}  we then focus on the explicit construction of differential graded 1-forms over the exceptional Jordan algebra $A = J_3(\mathbb{O})$, before covering the  more general setup for discrete exceptional geometries in Section \ref{Sec_disc_exc}. In this
paper we will not consider forms of degree higher than one.

\subsection{Review: associative differential 1-Forms}
	\label{universalforms}

Given an algebra $A$ (Jordan, associative, or otherwise), and  a bimodule $M$ defined over  $A$, then a derivation from $A$ into $M$ is a linear map $d:A\rightarrow M$ that satisfies:
\begin{align}
	d[ab] & = d[a]b + ad[b], & \forall a, b \in A\label{Leibniz}
\end{align}
Equation \eqref{Leibniz} is simply the Leibniz rule for the map `$d$' acting on $a, b \in A$.

The space of `universal' 1-forms defined over an \emph{associative} algebra $A$, denoted $\Omega_d^1A$, is the vector space generated as a left (or equivalently right) $A$-module by the symbols  $d[a]$, for $a\in A$, and where $d$ satisfies \eqref{Leibniz}. Associativity, together with the relation given in \eqref{Leibniz} ensures that a general element of $\Omega_d^1 A$ can be expressed as~\cite{landi1997}:
\begin{align}
	\omega = \sum ad[b],
\end{align}
where the sum is over elements $a,b\in A$. An explicit construction of $\Omega_d^1 A$ can be built over a unital, associative algebra $A$ by making use of its tensor algebra~\cite{landi1997}. First, the space $M=A\otimes A$ naturally forms a bimodule over $A$, with the following left and right actions:
\begin{align}
	\pi_L(a)(b\otimes c)&= ab\otimes c\in M\\
\pi_R(a)(b\otimes c)&= b\otimes ca\in M
\end{align}
$a\in A$, $b\otimes c\in M$. Next consider the submodule of $M$ given by
\begin{align}
	&ker(m:M\rightarrow A), & m(a\otimes b) =ab.\label{1forms}
\end{align}
We refer to this submodule as $\Omega^1_\Delta A$. Any element $\omega\in \Omega^1_\Delta A$ is of the form $\sum_i a_i\otimes b_i$ with $\sum_i a_ib_i= 0$. We are therefore free to write $\omega = \sum_i a_i(\mathbb{I}\otimes b_i - b_i\otimes \mathbb{I})$, from which we see that $\Omega^1_\Delta A$ is generated as a left module by elements of the form $a\otimes 1 - 1\otimes a$. An analogous argument from the right shows that these elements also generate $\Omega^1_\Delta A$ as a right module. Furthermore, if one  introduces the map $\Delta: A\rightarrow \Omega_\Delta^1 A$ given by\cite[Ch. 3]{Bourbaki}:
\begin{align}
	\Delta[a] =a\otimes e^0 - e^0\otimes a\label{deri}
\end{align}
for $a\in A$ and identity element  $e^0\in A$, then one notices that this map is linear, and acts as a derivation:
\begin{align}
	\Delta[ab] &= 	\Delta[a]b+	a\Delta[b], & &\forall a,b\in A. 
\end{align}
There is then an isomorphism of bimodules $\Omega_d^1 A\simeq  \Omega_\Delta^1 A$, with $ad[b]\leftrightarrow a(b\otimes 1 - 1\otimes b)$. By identifying $\Omega_\Delta^1 A$ with $\Omega_d^1 A$, the differential $d:A\rightarrow \Omega_d^1 A$ is given by~\cite{landi1997}:
\begin{align}
 d[a]&= \Delta[a], & a&\in A.
\end{align}
The reason that $\Omega_d^1A$ is known as the space of `universal' 1-forms, is that given any associative  $A$ bimodule $M$, with derivation $\delta:A\rightarrow M$, there exists a unique bimodule homomorphism $\phi:\Omega_d^1 A \rightarrow M$ such that  $\delta = \phi\circ d$, i.e. such that the following diagram commutes~\cite{landi1997}:
\begin{equation}
	\begin{tikzcd}
		A \arrow[d, "d"] \arrow[dr,  "\delta"] &  \\
		\Omega^1_d A \arrow[r, "\phi"] & M
	\end{tikzcd}
	\label{UPomega1}
\end{equation}
To see this, we need to show that the map: $\phi(d[a]):= \delta[a]$  extends consistently as   a well defined bimodule homomorphism to all of $\Omega_d^1 A$, and that it is unique. Since $\Omega_d^1 A$ is generated as a bimodule,  we extend $\phi$ as a bimodule homomorphism by defining:
\begin{align}
	\phi(\sum ad[b]c) = \sum a\delta[b]c
\end{align} 
We then  check that $\phi$ respects all of the relations on $\Omega_d^1 A$, and in particular Eq.~\eqref{Leibniz}. This can be seen directly:
\begin{align}
\phi(d[ab])&=\delta[ab]=a\delta[b]+\delta[a]b=\phi(ad[b]+d[a]b),& a,b&\in A
\end{align}
where we have used the fact that $\delta$ is a derivation in the second equality and that $\phi$ is defined as a bimodule homomorphism in the last equality.
Then, to show uniqueness,  suppose that $\psi$ is another bimodule homomorphism satisfying $\psi(d[a]) = \delta[a]$. Because $\Omega_d^1 A$  is generated by symbols $d[a]$ as both a  left and right module,  and $\phi$ and $\psi$ agree on these generators, they must be equal on all of $\Omega_d^1 A$. This ensures that $\phi$ is unique.

It is instructive to take a closer look at the explicit form that bimodule homomorphisms take in the finite dimensional, associative setting, and how these might be distinguished from module maps  in the nonassociative setting. In particular, for our explicit construction of $\Omega_d^1A$ we defined the derivation $\Delta:A\rightarrow A\otimes A$ as in equation \eqref{deri}. In general, this choice of derivation is not the only map from $A$ into $A\otimes A$ that is compatible with the Leibniz rule in \eqref{Leibniz}. The most general linear operator from $A$ into $A\otimes A$ is given by:
\begin{align}
\Delta(e^i) :=\Delta^i_{jk}(e^j\otimes e^k) \label{generaldir}
\end{align} 
where the $\Delta^i_{jk}$ are real or complex  valued coefficients, and the $e^i$ form a basis for $A$, with identity element $e^0$. The derivation given in \eqref{deri} corresponds to the case in which $\Delta^i_{i0} = -\Delta^i_{0i}=1$, for each $i$, and where all other coefficients are zero. If we express the  product on the coordinate algebra  $A$ in terms of its structure constants  $e^ie^j = f^{ij}_ke^k$, then the Leibniz rule given in \eqref{Leibniz} becomes:
\begin{align}
	f^{ij}_k\Delta_{mn}^k - f^{kj}_n\Delta_{mk}^i- f^{ik}_m\Delta_{kn}^j=0 \label{liebstruc}.
\end{align}
For a given set of structure constants, this equation can be used to constrain the $\Delta^i_{jk}$. In the finite dimensional, associative case, the choice $\Delta^i_{i0} = -\Delta^i_{0i}=1$ is generally not unique, due to module homomorphisms $\phi:A\otimes A\rightarrow A\otimes A$ of the form:
\begin{align}
	\phi(h\otimes k) = \sum (h u \otimes v k)\label{modmap},
\end{align}
where the sum is over the $u,v\in A$. The associativity of $A$ ensures that such transformations satisfy the standard properties of a bimodule homomorphism: $\sum a\phi(\omega)b = \sum \phi(a\omega b)$. The composition of any such bimodule map with a derivation $\Delta:A\rightarrow M=A\otimes A$ will always define a new derivation $\Delta_\phi=\phi\circ \Delta$. As such, for associative algebras one is generally able to define a whole family of operators $\Delta_\phi$ satisfying the Leibniz rule given in \eqref{Leibniz}. As we will see in section \ref{explicdga}, the story is somewhat more restrictive when considering derivations from the exceptional Jordan algebra into split Jordan bimodules, because maps of the form given in \eqref{modmap} will generally speaking no longer act as bimodule homomorphisms.

Finally, we end this section with a brief description of Connes' 1-forms in noncommutative spectral geometry. Given an associative spectral triple $T=(A,H,D)$, one is able to construct the space of universal 1-forms $\Omega_d^1 A$ over the coordinate algebra $A$. The Dirac operator then provides a map $\delta:A\rightarrow B(H)$ into the bounded operators on $H$, which act naturally as a bimodule over $A$:
\begin{align}
	\delta[a]:=[D,\pi(a)].\label{map-delta}
\end{align}
The fact that this provides a representation of `exact' universal 1-forms in the discrete, finite-dimensional cases of interest follows from the property that, in associative algebras, commutators act as derivations. This ensures that $\delta$ defines a module homomorphism $\pi:\Omega_d^1 A\rightarrow \Omega_D^1 A\in End(H)$, with $\pi(d[a]):=\delta[a]$.

\subsection{Exceptional Jordan 1-forms and Universality}

\label{explicdga}

In the previous section we explored derivations from associative algebras into associative bimodules. Now, consider  the exceptional Jordan algebra $A=J_3(\mathbb{O})$ and the split Jordan bimodule $M=A\otimes A$ defined over $A$. Our goal is to determine the structure of  derivations $\Delta:A\rightarrow A\otimes A$. Defining 
$\Delta(e^i) :=\Delta^i_{jk}(e^j\otimes e^k)$ as in Eq. \eqref{generaldir}, and imposing the Leibniz rule given in \eqref{liebstruc}, we find through an explicit  computation performed using Mathematica  that derivations are restricted to be of the form
\begin{align}
\Delta_\kappa[a] = \kappa(a\otimes e^0 - e^0\otimes a)
\end{align} 
 for $a\in J_3(\mathbb{O})$ and where $e^0\in J_3(\mathbb{O})$ is the identity element, and  $\kappa\in \mathbb{R}$ is a free parameter. It is no coincidence that up to a coefficient $\kappa$ there is only one available derivation from $A$ in $A\otimes A$.
 The freedom that exists stems from bimodule homomorphisms of the form $\phi_\kappa:A\otimes A\rightarrow A\otimes A$ with $\phi_\kappa(a\otimes b) := \kappa(a\otimes b)$, $\kappa\in \mathbb{R}$. 
Following Section~\eqref{sec_mod_maps} we know that all module homomorphisms from $M$ to itself must take this restricted form. 
 More general maps of the form given in Eq.~\eqref{modmap} no longer satisfy the properties of a bimodule homomorphism due to the nonassociativity of $A$. As a result, such maps are no longer available to generate a family of derivations $\Delta_\phi = \phi\circ \Delta$.  
 
We would like to define a notion of `split' universal differential 1-forms over $A=J_3(\mathbb{O})$. We begin in analogy with the associative setting by defining $\Omega_\Delta^1 A$ as the smallest submodule of $A\otimes A$ generated as a split Jordan bimodule by elements of the form:
\begin{align}
	\Delta[a] &= a\otimes e^0 - e^0\otimes a, & a&\in A.
\end{align} 
In the associative setting $\Omega_\Delta^1 A$ is generally smaller than $A\otimes A$, but it turns out for the Exceptional Jordan algebra that the elements $\Delta[a]$ generate the whole of $A\otimes A$ as a split Jordan bimodule. We include the proof, which is not difficult but which would otherwise break up the narative flow, in Appendix \ref{sec_jorda_identa}. 

Just as in the associative setting, the space of 1-forms $\Omega_\Delta^1 A$ is in fact universal in the following sense: Given any split Jordan bimodule of the form given in Eq.~\eqref{newmoduleform} $M =A\otimes V\otimes A$, defined over the exceptional Jordan algebra $A=J_3(\mathbb{O})$, with derivation $\delta:A\rightarrow M$, there exists a unique bimodule homomorphism $\phi:\Omega_\Delta^1 A \rightarrow M$ such that  $\delta = \phi\circ d$. In particular, from  section \ref{sec_mod_maps}, if $\phi$ is a bimodule homomorphism then it will take the form:
\begin{align}
	\phi(\Delta[a]) = a\otimes \Gamma^iv_i\otimes  e_0 - e_0\otimes \Gamma^iv_i\otimes a,
\end{align}
$a\in A$, and where the $\Gamma^i$ are real coefficients characterising the module map, with the $v_i$ forming a basis for $V$. It is clear that the coefficients $\Gamma^i$ define $\phi$ uniquely. Furthermore, it is quick to show that $\phi(\Delta[ab])=\delta[ab]=a\delta[b]+\delta[a]b=\phi(a\Delta[b]+\Delta[a]b)$, and therefore that $\phi$ respects the Leibniz relation of $\Delta$ on $\Omega_\Delta^1 A$. For this reason we  identify $\Omega_\Delta^1 A$ as the `split' universal 1-forms $\Omega_d^1 A$ over the exceptional Jordan algebra $A=J_3(\mathbb{O})$, with the differential $d:A\rightarrow \Omega_d^1 A$  given by:
\begin{align}
	d[a]&= \Delta[a], & a&\in A.
\end{align}
We point out again that  `universal' 1-forms over the exceptional Jordan algebra have already been defined in ~\cite{carotenuto2019,DV:2016}. Our definition of `split' universal 1-forms clearly doesn't  agree with this earlier notion of universal 1-forms. In particular, the space of universal 1-forms in \cite{carotenuto2019,DV:2016} is given by $\Omega_d^1 A =A\otimes Der^*(A)$, where $Der^*(A)$ is the 52 dimensional dual vector space to the space of inner derivations on $A$.  The distinction arises because we are looking at derivations  into split Jordan bimodules, while  \cite{carotenuto2019,DV:2016} considered maps into un-split Jordan bimodules. Notice that in the associative setting there is no distinction between split or un-split bimodules, meaning that this  differentiation isn't necessary.

\subsection{Discrete Geometry and Split  Universal 1-forms}
\label{Sec_disc_exc}

Having looked at the space of universal split Jordan 1-forms on a single copy of the exceptional Jordan algebra,   we now consider finite dimensional, discrete, exceptional Jordan algebras $A=\oplus_{1}^nJ_3(\mathbb{O})$, and split Jordan bimodules of the form 
$M = J_3(\mathbb{O})\otimes V\otimes J_3(\mathbb{O})$. We start with the `minimal' case in which each subspace of $V$ is given by $P^{ab}(V) = \mathbb{R}$ for each $a,b = 1,...,n$, i.e. $M=J_3(\mathbb{O})\otimes \mathbb{R}^{n^2}\otimes J_3(\mathbb{O})$. It will be useful to work in a basis for $A$:
\begin{align}
	e^{(a)j} := (0, \dots, e^j, \dots, 0) \in A,
\end{align}
where the $e^j$ on the RHS is at position `$a$', corresponding to a basis element of the $a^{th}$ $J_3(\mathbb{O})$ factor in $A$. Similary, we introduce the following basis elements for $M$
\begin{align}
	e^{i}\otimes^{ab} e^j := (0,...,e^{i}\otimes e^j,...,0)\in  M
\end{align}
where the $e^{i}\otimes e^j\in J_3(\mathbb{O})\otimes \mathbb{R}\otimes J_3(\mathbb{O})$ on the RHS are the basis elements which are isolated by the $P^{ab}$ projector on V. With this basis established the most general linear operator $\Delta:A\rightarrow M$ is of the form:
\begin{align}
	\Delta[e^{(a)i}] = \Delta^{(a)i}_{(bc)jk} e^j\otimes^{bc}e^k,
\end{align}
where the $\Delta^{(a)i}_{(bc)jk}$ are real coefficients. For $\Delta$ to act as a derivation it must further satisfy the Leibniz condition given in  Eq.\eqref{Leibniz}:  $\Delta[e^{(a)i}e^{(b)j}] = \Delta[e^{(a)i}]e^{(b)j}+e^{(a)i}\Delta[e^{(b)j}]$. This implies the following relations, which must be satisfied by the $\Delta^{(a)i}_{(bc)jk}$ coefficients:
\begin{align}
	\delta^{(ab)}f_k^{ij}\Delta^{(b)k}_{(cd)mn} - \delta^{(bd)}f_n^kj \Delta^{(a)i}_{(cd)mk} - \delta^{(ac)}f_m^{ik}\Delta_{(cd)kn}^{(b)j}=0,\label{discex1}
\end{align}
where the $f^{ij}_k$ are the structure constants that define the product on each copy of $J_3(\mathbb{O})$, and where Einstein's summation convention is not being used with  bracketed indices. This equation is really just  Eq.~\eqref{liebstruc} expressed in our basis for $A$ and $M$. While this expression looks cumbersome, it fortunately decomposes into a number of manageable  cases. In particular, if we consider the case in which $d\neq a=b\neq c$, then Eq.~\eqref{discex1}  becomes:
\begin{align}
f_k^{ij}\Delta^{(b)k}_{(cd)mn} =0,
\end{align}
which tells us immediately that:
\begin{align}
\Delta^{(a)}_{(bc)} &=0, & \text{for }&a\neq b\text{ and }a\neq c,
\end{align}
where $\Delta^{(a)}_{(bc)}$ correspond to the components of $\Delta$ that map from basis elements $e^{(a)i}$ into the subspace of $M=J_3(\mathbb{O})\otimes V\otimes J_3(\mathbb{O})$ isolated by projector  $P^{bc}$  on $V$. We therefore only need to worry about 
the components of $\Delta$ that map from basis elements $e^{(a)i}$ into the subspace of $M$ isolated by projectors of the form $P^{ab}$ and $P^{ba}$  on $V$.  We begin by isolating  the `diagonal' components of $\Delta$ that map basis elements $e^{(a)i}$ into the subspace of $M$ isolated by $P^{aa}$, by setting $a=b=c=d$ in Eq.~\eqref{discex1}. This yields:
\begin{align}
	f_k^{ij}\Delta^{(a)k}_{(aa)mn} - f_n^kj \Delta^{(a)i}_{(aa)mk} - f_m^{ik}\Delta_{(aa)kn}^{(a)j}=0,
\end{align}
which is the exact same expression given in Eq.~\eqref{liebstruc}, which is  satisfied by a derivation from a single copy of $J_3(\mathbb{O})$  into a split Jordan bimodule $J_3(\mathbb{O})\otimes J_3(\mathbb{O})$. Following section \ref{explicdga}, we therefore have:
\begin{align}
	\Delta^{(a)}_{(aa)}[e^{(a)i}] &= \kappa_{(aa)}(e^i\otimes^{aa} e^0 - e^0\otimes^{aa} e^i), & \kappa_{(aa)}\in \mathbb{R}.\label{aadisc}
\end{align}
We can similarly isolate the components $\Delta^{(a)}_{(ab)}$ and $\Delta^{(a)}_{(ba)}$ of $\Delta$ by considering the cases $a=c\neq d=b$,  $a=b=c\neq d$, and $a=b=d\neq c$ in Eq.~\eqref{discex1}. This yields the following three  coupled equations:
\begin{align}
	f^{kj}_{n}\Delta^{(e)i}_{(ef)mk}+f^{ik}_m\Delta^{(f)j}_{(ef)kn}&=0,\\
		f^{ij}_{k}\Delta^{(e)k}_{(ef)mn}-f^{ik}_m\Delta^{(e)j}_{(ef)kn}&=0,\\
			f^{ij}_{k}\Delta^{(f)k}_{(ef)mn}-f^{kj}_n\Delta^{(f)i}_{(ef)mk}&=0.
\end{align}
 Solving these equations through an explicit  computation performed using Mathematica, we find that derivations are restricted to be of the form:
\begin{align}
	\Delta^{(c)}_{(ab)}[e^{(c)i}] &= \kappa_{(ab)}(\delta^{(ca)}e^i\otimes^{ab} e^0-\delta^{(bc)}e^0\otimes^{ab} e^i), & \kappa_{ab}&\in\mathbb{R}.\label{abdisc}
\end{align}
Combining Eq.~\eqref{aadisc} and \eqref{abdisc}, we therefore have in general
\begin{align}
		\Delta_\kappa[e^{(a)i}] = \sum_c  (\kappa_{(ac)} e^i\otimes^{ac}  e^0 - \kappa_{(ca)}e^0\otimes^{ca} e^i ).
\end{align}
We see that up to the coefficients $\kappa_{(ab)}$ there is a single unique form that derivations can take from $A$ into $M$. Furthermore, from Eq.~\eqref{discretmodorpph} we see that the freedom that does exist stems from  bimodule morphisms of the form $	\phi_\kappa(a\otimes^{ab}\otimes b) = 	\kappa_{ab}(a\otimes^{ab} b)$.  For this reason we define $\Omega_\Delta^1 A$ as the smallest submodule of $M$ generated as a split Jordan bimodule by elements of the form:
\begin{align}
	\Delta[e^{(a)i}] = \sum_c  (e^i\otimes^{ac}  e^0 - e^0\otimes^{ca} e^i ).
\end{align}
Once again, following an argument analogous to that shown in Appendix Section \ref{sec_jorda_identa}, it is quick to show that $\Omega_\Delta^1 A=J_3(\mathbb{O})\otimes \mathbb{R}^{n^2}\otimes J_3(\mathbb{O})$. That is, the elements $\Delta[a]$ generate all of $\Omega_\Delta^1 A=J_3(\mathbb{O})\otimes \mathbb{R}^{n^2}\otimes J_3(\mathbb{O})$ as a split Jordan bimodule.

Finally, it turns out that the space of 1-forms $\Omega_\Delta^1 A$ is universal in the following sense: Given any split Jordan bimodule of the form $M =J_3(\mathbb{O})\otimes V\otimes J_3(\mathbb{O})$, defined over the exceptional Jordan algebra $A=\oplus_1^nJ_3(\mathbb{O})$,
with left and right actions defined as in \eqref{mult11} and \eqref{mult2}, and  with derivation $\delta:A\rightarrow M$, then there exists a unique bimodule homomorphism $\phi:\Omega_\Delta^1 A \rightarrow M$ such that  $\delta = \phi\circ \Delta$. In particular, following section \ref{sec_mod_maps}, if $\phi$ is a module homomorphism then it will be of the form:
\begin{align}
\phi(\Delta[e^{(a)i}]) =\sum_c  (e^i\otimes P^{ac}(\Gamma^iv_i) \otimes  e^0 - e^0\otimes P^{ca}(\Gamma^iv_i) \otimes e^i )\label{transformD}
\end{align}
$a\in A$, and where the $\Gamma^i$ are real coefficients characterising the module map, with the $v_i$ forming a basis for $V$. It is clear that the coefficients $\Gamma^i$ constrained by the Projection operators $P^{ab}$ define $\phi$ uniquely. Furthermore, it is quick to show that $\phi(\Delta[ab])=\delta[ab]=a\delta[b]+\delta[a]b=\phi(a\Delta[b]+\Delta[a]b)$, and therefore that $\phi$ respects the Lebniz relation of $\Delta$ on $\Omega_\Delta^1 A$. For this reason we  identify $\Omega_\Delta^1 A$ as the `split' Universal 1-forms $\Omega_d^1 A$ over the exceptional Jordan algebra $A=\oplus^n_1J_3(\mathbb{O})$, with the differential $d:A\rightarrow \Omega_d^1 A$  given by:
\begin{align}
	d[a]&= \Delta[a], & a&\in A.
\end{align}

\subsection{Dirac Operators}
\label{dirac_operator}
We have now established the context and tool-set required to fully appreciate the rational behind  our example 2-point geometry constructed in Section \ref{example_geometry}. The Dirac operator in Eq.~\eqref{Dirac} provides a derivation $\delta:A\rightarrow \Omega_D^1 A\subseteq End(H)$ with $\delta(a)=[D,\pi(a)]$, $a\in A$, just as in Eq.~\eqref{map-delta}:
\begin{align}
[D,\pi(e^{(a)i})] &=\kappa\begin{pmatrix}
	0 & \delta^{a2} e_0\otimes^{12} e_i^\ast - \delta^{a1}e_i\otimes^{12} e_0^\ast \\
	\delta^{a1}e_0\otimes^{21} e_i^\ast  -\delta^{a2}e_i\otimes^{21} e_0^\ast &0
\end{pmatrix},
\end{align}
$e^{(a)i}\in A = J_3(\mathbb{O})\oplus J_3(\mathbb{O})$. The elements $[D,a]$ then generate  $\Omega_D^1 A$ as a split Jordan bimodule under the left and right actions given in Eqs.~\eqref{action11} and~\eqref{action22}. The derivation $\delta$ and split Jordan bimodule structure of $\Omega_D^1 A$ defines a bimodule homomorphism $\pi:\Omega_d^1 A = J_3(\mathbb{O})\otimes \mathbb{R}^4\otimes J_3(\mathbb{O})\rightarrow \Omega_D^1 A = J_3(\mathbb{O})\otimes \mathbb{R}^2\otimes J_3(\mathbb{O})$,  with $\pi(d[e^{(a)i}]) := [D,\pi(e^{(a)i})]$. Comparing with  Eq.~\eqref{transformD}, we see that the `diagonal' elements $J_3(\mathbb{O})\otimes^{aa}J_3(\mathbb{O})\subseteq \Omega_d^1 A$ sit in the kernel of $\pi$, while $\kappa$ parametrizes the map from the `off-diagonal' elements. In this light, we see that Hermiticity, together with the restricted form that bimodule homomorphisms between split Jordan bimodules take places severe constraint on the form of $D$. Our assumed off-diagonal form for $D$ is furthermore compatible with a grading condition $\{D,\gamma\}=0$ that is usually imposed when constructing gauge theories~\cite{Chamseddine:2007oz},  were $\gamma = diag(S_{e_0},-S_{e_0})$ is a  $\mathbb{Z}_2$ grading on $H$.

\section{Discussion}
\label{sec_discussion}
Nonassociative spectral geometries are difficult to build because modules play a central role in their construction,  and the representation theory for nonassociative algebras is
significantly more challenging, more fragmented, and less well developed than for associative algebras.  To build a spectral triple $T=(A,H,D)$,  one needs to construct a Hilbert space $H$ which acts as a module over the coordinate algebra $A$, and one needs to construct forms $\Omega_D A = \oplus_i \Omega^i_D A$, which act as  bimodules over $A$, and which are compatible with the notion that the Dirac operator $D$ should satisfy the properties of a first order derivative operator.

In previous work~\cite{Besnard_2022,Farnsworth_2020,Boyle:2020} we established a framework for constructing spectral geometries based on special Jordan algebras. Special Jordan algebras play an important role in the extension from associative to nonassociative spectral geometry because they have clear physical applications~\cite{Boyle:2020}, and because their representation theory is well understood in comparison to  most other nonassociative algebras. Importantly, for special Jordan algebras one has access to both  `associative' and `multiplicative' representations~\cite{Besnard_2022}. \emph{If}, when developing a special Jordan spectral triple $T=(A,H,D)$, one begins with an `associative' representation $\pi:A\rightarrow End(H)$ (see Sec.~\ref{sec_mod_gen} for definitions), then this representation has  a natural extension to 1-forms  $\pi:\Omega_d^1A\rightarrow  End(H)$ such that the product on  zero and one forms is given as in Eqs.~\eqref{formsproductspecial0} and \eqref{formsproductspecial}:
\begin{align}
	\pi(a)\star\pi(b) &=\pi(a\circ b)= \frac{1}{2}\{\pi(a), \pi(b)\}\label{productforms0}\\
 	\pi(a)\star\pi(d[b]) &=	\pi(a\circ d[b])= \frac{1}{2}\{\pi(a),[D, \pi(b)]\}\label{productforms}
 \end{align} 
$a,b\in A$. For special Jordan geometries it is this `$\star$' product   that the Leibniz rule given in Eq.~\eqref{leib} is  established with respect to~\cite{Besnard_2022}. This generalises  the situation in associative spectral geometry in which the associative representation of $A$ on $H$ is extended to $A\oplus \Omega_d^1 A$, such that  the `$\star$' product between Connes' zero and one forms is given by the operator product.

The significance of Eq.~\eqref{productforms}  is that it clearly showcases the approach that should be taken in generalising from the associative setting.  Namely,  one has in general $\pi(\omega\omega')\neq \pi(\omega)\pi(\omega')$ for $\omega,\omega'\in A\oplus \Omega_d^1 A$. While working within the endomorphisms of $H$, one should use a product $\pi(\omega)\star\pi(\omega') := \pi(\omega\omega')$ that respects the algebraic structure of the  algebra being represented. 
The challenge is defining an appropriate bimodule structure such that the product `$\star$' remains compatible with the symmetries of the algebra representation on $H$, and such that the Dirac operator $D$ acts as a derivation with respect to `$\star$'. These requirements are fairly strict, and it is  not immediately clear  beyond the associative and special Jordan settings that it should even be possible to find an appropriate pair $(D,\star)$ for a given set of nonassociative order zero data $(A,H)$.
Exceptional Jordan algebras provide an ideal test bed for constructing proof-of-principle examples beyond the associative setting. This is because a well-defined trace form exists for these algebras ensuring that one is able to construct Hilbert spaces,
 and a theory of Jordan bimodules is already well established~\cite{jacob1968}.
 
 Exceptional Jordan algebras lack associative representations, but they do admit multiplicative ones. Since all exceptional Jordan bimodules are free~\cite{DV:2016}, constructing the order-zero data $(A,H)$ for an exceptional Jordan spectral triple is relatively straightforward, as there is little room for choice.
  However, if one next takes the seemingly natural approach of  requiring the space of 1-forms $\Omega_D^1 A$ over a discrete exceptional Jordan algebra $A=\oplus^n_1 J_3(\mathbb{O})$ to also act as a standard, non-split Jordan bimodule,  then  one seemingly arrives at an impasse.
   Specifically, one appears restricted to `derivation' based 1-forms~\cite{carotenuto2019,DV:2016}, defined on the inner derivations, which act `internally' within each factor of $A$.
   What is actually needed,  is an `external' derivation  into a  a bimodule that couples the  distinct points of the geometry. This motivates the adoption of split Jordan bimodules, which have distinct left and right actions.  
 For the curious reader, it is instructive to attempt replacing the space of 1-forms in our example two-point geometry (Sec.~\ref{example_geometry}) with a non-split bimodule to see where that approach fails.
 
The associative properties of exceptional Jordan algebras and their representation on split Jordan bimodules severely restrict the form that Dirac operators take in exceptional Jordan spectral  geometries. In particular, we saw that Hermiticity, compatiblility with a $\mathbb{Z}_2$ grading on $H$, and the form of bimodule homomorphisms restricted the Dirac operator  in our 2-point example geometry (Sec.~\ref{example_geometry}) to a single free parameter. This is in deep contrast to  both the  associative  and special Jordan settings~\cite{Besnard_2022}, where Dirac operators   are far less restricted. The restriction that appears in the exceptional Jordan setting is exciting, because it shows, as anticipated, that novel restrictions do indeed arise in the nonassociative setting, which from the perspective of model building are completely unexplored.

While we have shown that it is indeed possible to construct non-trivial and  self consistent  geometries, as well as  established an  approach  that is readily generalisable, a key question remains: How unique was our choice of selecting split Jordan bimodules? Consider again for instance,   the left and right actions defined on split Jordan bimodules given in Eqs.~\eqref{mult11} and \eqref{mult2}. These can  be combined to construct a new `symmetrized' action on split Jordan modules:
\begin{align}
	\pi_S(a)(h\otimes v\otimes k) & = \sum_{ij}^n \frac{1}{2}\left(a_i v \otimes P^{ij}(v)\otimes k+  v \otimes P^{ij}(v)\otimes a_jk\right) 
\end{align}
where $a = (a_1,...,a_n)\in A=\oplus^n_1 J_3(\mathbb{O})$, and  $P^{ij}$ projects onto the subspace $P^{ij}(V) = V^{ij}$ of $M = J_3(O)\otimes V\otimes J_3(O)$. While this new action is symmetric it no longer satisfies the multiplicative properties of Jordan modules given in Eqs.~\eqref{mult1} and\eqref{jordact},  meaning that this new symmetrized action does not correspond to the traditional definition of a Jordan bimodule. Nonetheless,  this action is compatible with the symmetries of $A$, and if the analysis of Sec.~\ref{explicdga} were to be repeated, our map $\Delta: A\rightarrow M$ would once again be found to act as a derivation. In this case, however, one would no longer find that $\Omega_\Delta^1A$ is identified with all of $M$, but rather with an anti-hermitian submodule.
 It appears that a  richer landscape of self consistent nonassociative spectral geometries
   may exist than  initially anticipated, arising from a potential variety of self-consistent bimodule structures, which one seemingly does not have any analog of in the purely  associative setting. 
   
 
An exploration of the full landscape of potential  nonassociative spectral geometries, and indeed a complete theory of nonassociative bimodules,
 lies outside the scope of the current  paper.
 Our analysis, also focuses specifically on the order zero and order one data of  spectral triples, leaving the exploration of  full differential graded calculi for future work. 
  Here we establish, to first order, the first   self-consistent examples of  discrete nonassociative geometries with non-trivial metric data, and without  associative envelope,  associative representations, or twisting from the associative setting.
We note, that the split universal 1-forms over exceptional Jordan algebras introduced in this paper do not replace the universal forms studied elsewhere~\cite{carotenuto2019}. Rather, they complement them, broadening the possibilities for model building. The exceptional Jordan algebra has been highlighted by many~\cite{boylef4,todorov,tod2,Bhatt_2022,Patel_2023,Krasnov_2021}
 as a potential key to understanding a number of otherwise unexplained features of the Standard Model of Particle Physics, including the origin of particle generations. The approach developed here, in conjunction with~\cite{Farnsworth:2013nza,carotenuto2019,Besnard_2022,Farnsworth_2020,Boyle:2020,Boyle:2014wba}, establishes a  rigorous foundation and novel avenues for such investigations, opening  in particular,  new possibilities for constructing self-consistent dynamical models, including those based on the spectral action~\cite{Chamseddine_1997}.


\section*{Acknowledgements}
I would like to thank Felix Finster, Claudio Paganini, and  Axel Kleinschmidt  for their support during the writing  of this work. I would also like to thank Keegan Flood and Tejinder Singh for useful discussions and  suggestions for improving  this text.

\appendix

\section{The Exceptional Jordan Algebra}

\label{sec_jorda}

A real Jordan algebra is a vector space $A$ over $\mathbb{R}$ equipped with a bilinear, commutative product $\circ$ satisfying the Jordan identity:
\begin{align}
	(a^2\circ b)\circ a = a^2\circ (b\circ a)
\end{align}
$\forall a,b\in A$. Note that while Jordan algebras are in general not associative, every Jordan algebra is power
associative, i.e. $a^n$ has an unambiguous meaning for all $n\in\mathbb{N}$. Furthermore, using the Jordan Identity it is possible to show that $(a^m\circ b)\circ a^n=a^m\circ ( b\circ a^n)$ for all $n,m\in\mathbb{N}$~\cite{jacob1968}. A Jordan algebra is said to be `special' if it can be constructed from an associative algebra. Specifically, let $A$ be an associative algebra equipped with the product
\begin{align}
a \circ b = \frac{1}{2}(ab + ba)	
\end{align}
and let $A^+$ be a real subspace of $A$ stable under $\circ$ (an important example is when $A$ is a $*$-algebra and
$A^+$ is the space of selfadjoint elements). Then $(A^+, \circ)$ is a Jordan algebra. Special Jordan algebras are  isomorphic to algebras of this kind. All other Jordan algebras are called exceptional. 


A Euclidean  Jordan algebra has the additional property that $a_1^2+a_2^2+...a_n^2 = 0$ implies $a_1 =a_2 = ... = a_n = 0$ for $a_i\in A$. These are the algebras identified by Jordan as suitable for describing quantum mechanical observables because they naturally capture the properties of symmetric operators with real eigenvalues.
In this paper we focus on a particular Euclidean Jordan algebra known as the exceptional Jordan algebra. \emph{The} exceptional Jordan algebra, otherwise known as the Albert algebra $J_3(\mathbb{O})$ is the algebra of  Hermitian $n \times n$ matrices, with octonionic entries.  When working with the Albert algebra it is useful to introduce two distinct  bases. The first is given by:
	\setlength{\arraycolsep}{1.5pt}
\begin{align}
	e^1 &= \begin{pmatrix}
		\theta_0&0&0\\
		0&0&0\\
		0&0&0
	\end{pmatrix}, &
	e^{10} &= \begin{pmatrix}
		0&0&0\\
		0&\theta_0&0\\
		0&0&0
	\end{pmatrix}, & 
	e^{19} &= \begin{pmatrix}
		0&0&0\\
		0&0&0\\
		0&0&\theta_0
	\end{pmatrix},\nonumber\\
	e^{i+2} &=\begin{pmatrix}
		0&0&0\\
		0&0&\theta_{i}\\
		0&\theta_{i}^*&0
	\end{pmatrix}, &
	e^{i+11} &=\begin{pmatrix}
		0&0&\theta_{i}^*\\
		0&0&0\\
		\theta_{i}&0&0
	\end{pmatrix}, & 
	e^{i+20}  &=\begin{pmatrix}
		0&\theta_{i}&0\\
		\theta_{i}^*&0&0\\
		0&0&0
	\end{pmatrix},\nonumber
\end{align}
where the $\theta_i$, $i=0,...,7$ form a basis for the octonions and where $*$ denotes the involution on $\mathbb{O}$. In this basis the identity element is given by the linear combination $e^0=e^1+e^{10}+e^{19}$. The  structure constants $T^{ij}_k$ defined in terms of the product $e^{i}\circ e^j = T^{ij}_ke^k$ are extremely sparse and easy to work out by hand when expressed in terms of the product on the octonions. This basis also has three fold symmetries that make is well suited for explicit calculations. This is the basis we made use of
in sections \ref{explicdga} and \ref{Sec_disc_exc} when using mathematica to determine the form of derivatoion elements $\Delta$ from an exceptional Jordan coordinate algebra into a split jordan bimodule.

For any finite dimensional Euclidean Jordan algebra, a basis $\{\sigma^0,\sigma^1,...,\sigma^n\}$ can be chosen such
that $\sigma^i\circ\sigma^j = \delta^{ij}\sigma^0 +T^{ij}_k\sigma^k$, where $\sigma^0$ is the identity element~\cite{townsandreview}. 
 In the case of the exceptional Jordan algebra, one can define  $\sigma^0=e^1+e^{10}+e^{19}$, $\sigma^1 = \sqrt{\frac{3}{2}}(e^1-e^{10})$, $\sigma^2 = \frac{1}{\sqrt{2}}(e^1+e^{10}-2e^{19})$, $\sigma^{i=3,...,10}=\sqrt{\frac{3}{2}} e^{i-1}$, $\sigma^{i=11,...,18} = \sqrt{\frac{3}{2}}e^i$, and $\sigma^{i=19,...,26} = \sqrt{\frac{3}{2}}e^{i+1}$. While this `$\sigma$' basis is not as symmetric as the `$e$' basis, and therefore slightly less manageable  for certain kinds of brute force computations, it is more useful when defining trace forms and pure states on $J_3(\mathbb{O})$.

\section{Pure states on $A = J_3(\mathbb{O})$}
\label{sec_state}

Given a  Euclidean Jordan algebra $A$, with basis defined
as at the end of Appendix Section \ref{sec_jorda}, such that $\sigma^i\circ\sigma^j = \delta^{ij}\sigma^0 +T^{ij}_ke^k$, a trace form can be defined as~\cite{townsandreview}:
\begin{align}
	Tr[\sigma^0] &= \nu, & Tr[\sigma^i]&=0,\label{traceform}
\end{align}
 The introduction of  such a trace form is equivalent to introducing a positive definite inner product
\begin{align}
	\langle\sigma^i| \sigma^j\rangle:=\frac{1}{\nu}Tr[\sigma^i\circ \sigma^j] = \delta^{ij},\label{innerprod}
\end{align}
where the indices $i,j$ range over the full basis, including the identity element $\sigma^0$. This Hilbert space structure on $A$ allows states on the algebra to be defined in the usual way as positive linear maps from the algebra to the field over which the algebra is defined:
\begin{align}
	\rho_a(b) &= 	\langle a|b\rangle, & a,b&\in A\label{states1}
\end{align}
for $a,b\in A$.  A state $\rho_a$ is defined in terms of an element $a\in A$, normalized such that $\rho(e_0)=1$. Any element of a Euclidean Jordan algebra can be expressed as a sum of what are known as `primitive idempotents'~\cite{jordanneumannwigner}, and it  is this decomposition that allows states on an algebra to be expressed as convex sums of `pure' states. An idempotent of a Euclidean Jordan algebra $A$ is an element $p\in A$  satisfying $p^2=p$. Any two  idempotents $p_1$ and $p_2$ that are orthogonal with respect to the inner product satisfy~\cite{townsandreview}
\begin{align}
	p^1 \circ p^2 = 0. \label{orthog}
\end{align}
An idempotent $E$ that cannot be expressed as the sum of two orthogonal idempotents is said to be primitive. The maximal number of mutually orthogonal primitive idempotents is the
degree of the Jordan algebra `$\nu$', and any such set provides a decomposition of unity:
\begin{align}
	\sum_{\alpha=1}^\nu E^\alpha =\sigma^0 
\end{align}
If the trace form is normalized as in \eqref{traceform} then the degree is the integer $\nu$. The primite idempotents represent pure states. A mixed state is represented by:
\begin{align}
	a &= \sum_{\alpha=1}^\nu a_\alpha E^\alpha, & \sum_\alpha^\nu a_\alpha &= 1, & a_\alpha\ge 0. 
\end{align}
In the case of the exceptional Jordan algebra, the trace form can be expressed in terms of the usual trace on operators $Tr[a]\propto Tr[S_a]$, where $S_a$ is the operator that details the action of $a\in A$ on elements of the algebra itself $S_ab = ab$~\cite{baez}. It turns out that the automorphisms of the exceptional Jordan algebra can be used to  `diagonalized' $S_a$ for any $a\in J_3(\mathbb{O})$.  The cyclicity of the  operator trace ensures invariance under the automorphisms of the algebra. This means that up to automorphism every `primive idempotent' in $J_3(\mathbb{O})$ can be expressed in  the form~\cite{baez}:
\begin{align}
	p^0 &= 0,&	p^1 &= \nu e^1, & p^2&=\nu  e^{10},& p^3&=\nu  e^{19}.
\end{align}
Such   primitive idempotents transform under the automorphisms to take the more general form:
\begin{align}
	p = \begin{pmatrix}
		x\\y\\z
	\end{pmatrix}\begin{pmatrix}
	x^* & y^* & z^* 
\end{pmatrix} & = \begin{pmatrix}
xx^* & xy^* & xz^*\\
yx^* & yy^* & yz^*\\
zx^* & zy^* & zz^*\\
\end{pmatrix}
\end{align} 
where the $(x,y,z)\in\mathbb{O}$ satisfy
\begin{align}
(xy)z &= x(yz), & ||x||^2+||y||^2+||z||^2 =\nu.
\end{align}
The primitive idempotents map out what is known as the `octonionic' projective plane $\mathbb{OP}^2$, which is a 16 dimensional manifold~\cite{baez}.

\section{Connes' Distance in Exceptional 2-point spaces}
\label{sec_2pointdistance}

In Section \ref{example_geometry} we introduced a 2-point exceptional Jordan geometry coordinatized by the algebra $A = J_3(\mathbb{O})\oplus J_3(\mathbb{O})$, represented on $H= J_3(\mathbb{O})\otimes \mathbb{R}^2$, with the action on $H$ given by the two point representation in Eq.~\eqref{Exceptional_Rep}. The Hilbert space is equipped with a natural inner product defined by:
\begin{align}
	\langle h| v\rangle = \frac{1}{2\nu}(Tr[h_1v_1] + Tr[h_2,v_2]),
\end{align}
for $v=(h_1,h_2),v=(v_1,v_2)\in H$, and the trace is defined as Eq.~\eqref{traceform}. The algebra representation given in \eqref{Exceptional_Rep} is symmetric with respect to this inner product. The states on $A$  can  be expressed as in Eq.~\eqref{states1}:
\begin{align}
	\rho_a(b) 
	&= \sum_{\alpha=1}^3 \frac{1}{2\nu}(a_{1,\alpha} Tr[{E^\alpha_1b_1}] + a_{2,\alpha} Tr[{E^\alpha_2b_2}])
\end{align}
$a=(a_1,a_2),b=(b_1,b_2)\in A$
where the $E_a^i$ are primitive idempotents defined on the $a=1,2$ factor of the algebra. The states are defined such that $\sum_{i\alpha} a_{\alpha,i} = 1$. Notice that the primitive idemponents $E_a^1$ and $E_a^2$ transform amongst themselves themselves under the automorphisms of the algebra. Without loss of generality we can  therefore always choose a basis such that any pure state takes the form:
\begin{align}
	x&= 2\nu\begin{pmatrix}
		e^1&0\\
		0&0
	\end{pmatrix}, & y&=2\nu\begin{pmatrix}
	0&0\\
	0&e^1
\end{pmatrix}.
\end{align} 
Next, we would like to define the distance between the two points $x$ and $y$,   given in terms of Connes Distance function 
\begin{align}
	d(x,y) = sup\{|\rho_x(a) - \rho_y(a)|: ||[D,\pi(a)]||\le 1\},
\end{align}
$a\in A$. The overlap between pure states and algebra element will be maximaized for an algebra element of the form: 
\begin{align}
	a = \begin{pmatrix}
		\alpha e_1 &0\\
		0& \beta e_1
	\end{pmatrix},
\end{align}
which yields $|\rho_x(a) - \rho_y(a)| = |\alpha-\beta|$. 
Making use of the Dirac operator given in Eq.~\eqref{Dirac}, we then calculate:
\begin{align}
[D,\pi(a)] = \kappa \begin{pmatrix}
	0&\beta(e_0\otimes e_1)-\alpha(e_1\otimes e_0)\\
	\alpha(e_0\otimes e_1)-\beta(e_1\otimes e_0)
\end{pmatrix}
\end{align}
which tells us that $||[D,\pi(a)]|| = max\{\kappa\alpha,\kappa\beta,\kappa(\alpha-\beta)\}$. This yields:
\begin{align}
	d(x,y) = \frac{1}{\kappa}.
\end{align}
We  see that the parameter $\kappa$ sets the distance between the two points.

\section{Identifying  $\Omega_\Delta^1 A$}

\label{sec_jorda_identa} 
In Section \ref{explicdga} we introduced the map $\Delta:A\rightarrow \Omega_\Delta^1 A$,  $\Delta[a] = a\otimes e_0 - e_0\otimes a$, for the exceptional Jordan algebra $A=J_3(\mathbb{O})$. We defined $\Omega_\Delta^1 A$ as the smallest subspace of $A\otimes A$ generated as 
a split Jordan bimodule by elements $\Delta[a]$, $a\in A$. It turns out, however, that $\Omega_\Delta^1 A$ can be identified with all  of $A\otimes A$. To see this, consider the following element:
\begin{align}
	-e^1\Delta[e^j] &= -(S_{e^1}e^j)\otimes e^0 + (S_{e^1}e^0)\otimes e^j\nonumber\\
	&=e^1\otimes e^j \label{1term}
\end{align} 
where we are making  use of the basis given in Section \ref{sec_jorda}, and where $2\le j\le 10$, or $j = 19$. Note that the brackets on the right hand side of the above expression are not really necessary, but are included for clarity. In the above expression we have specifically chosen the algebra elements $e_1$ and $e_j$ to ensure that we obtain a single term rather than a sum of terms. Next, notice that we can construct inner derivation elements acting independently on the left side of elements  $\sum a\otimes b\in A\otimes A$ as follows:
\begin{align}
	e(c(a\otimes b)) - 	c(e(a\otimes b)) & = ([S_e,S_c]a)\otimes b\nonumber\\
	&= \delta_{e,c}a\otimes b, \label{derleft}
\end{align}
$a,b,c,e\in A$, and where again the brackets on the right hand side of the above expression are superfluous. Similarly we can construct inner derivation elements that act independently on the right side  of elements $\sum a\otimes b\in A\otimes A$. As these inner derivation elements generate  $F_4$, this means that we can use the left and right actions of the algebra itself to independently rotate both sides of the tensor product in Eq.\eqref{1term} under the $F_4$ symmetry.

It turns out, that the primitive idempotentents $e^1,e^{10},$ and $e^{19}$ can be transformed into one another under the action of the $F_4$ symmetry group~\cite{baez}. This means, in particular that we can transform the element given in Eq.~\eqref{1term} into the elements $e^{10}\otimes e^j$ and $e^{19}\otimes e^j$. These elements can then be used to construct:
\begin{align}
	e^0\otimes e^j = (e^1+e^{10}+e^{19})\otimes e^j
\end{align}
Multiplying from the left by an algebra $a\in A$, we can then construct any element of the form $a\otimes e_j \in A\otimes A$. Notice, that we started the above argument with an element of the form $e^1\Delta[e^j]$ where the $e^j$ were those elements for which $e^1\circ e^j=0$. We could, however have started with $e^{10}\Delta[e^j]$  with those elements $e^j$ for which $e^{10}\circ e^j=0$, or with $e^{19}d[e^j]$  with those elements $e^j$ for which $e^{19}\circ e^j=0$. As these choices span the full basis of $e^j\in A$, we can therefore construct any element $a\otimes b\in A\otimes A$. We therefore have $\Omega_\Delta^1 A=A\otimes A$.

As an aside, one might wonder why a similar argument doesn't work for finite dimensional associative algebras. The problem in the associative setting is that derivations take the form of commutators as in Eq.~\eqref{assocdir}. It is easy to convince oneself that it is not possible to make use of the right and left actions of the algebra on $A\otimes A$ to construct derivation elements which act independently on the two sides of the tensor product as in Eq.~\eqref{derleft}.




\begin{thebibliography}{10}
	
	
	\bibitem{marco1} D. Denicola, M. Marcolli, and A. Z. al-Yasry, 
	``Spin foams and noncommutative geometry," 
	\textit{Class. Quantum Grav.} \textbf{27}, 205025 (2010).  
	\href{https://doi.org/10.1088/0264-9381/27/20/205025}{doi:10.1088/0264-9381/27/20/205025}.
	
	
	
\bibitem{AlfsenShultz1998} E. Alfsen and F. Shultz,  
``On orientation and dynamics in operator algebras. Part I,"  
\textit{Commun. Math. Phys.} \textbf{194}, 87–108 (1998). 
	\href{	https://doi.org/10.1007/s002200050350}{	doi:10.1007/s002200050350}.

	
\bibitem{Farnsworth_2020} S. Farnsworth,  
``The geometry of physical observables,"  
\textit{J. Math. Phys.} \textbf{61}, 102101 (2020).  
\href{https://doi.org/10.1063/5.0021707}{doi:10.1063/5.0021707}.

	
	\bibitem{Connes:1994kx} A. Connes,  
	``Noncommutative Geometry,"
	(Academic Press, 1994).  
	ISBN: 978-012-1858-605.
	

\bibitem{Connes:reconstruction} A. Connes,  
``On the spectral characterization of manifolds,"  
\textit{J. Noncommut. Geom.} \textbf{7}(1), 1–82 (2013).  
\href{https://doi.org/10.4171/JNCG/108}{doi:10.4171/JNCG/108}.

	
\bibitem{Connes:2008kx} A. Connes and M. Marcolli,  
``Noncommutative Geometry, Quantum Fields and Motives," 
(American Mathematical Society and Hindustan Book Agency, 2008). ISBN: 978-147-0450-458.
	


\bibitem{Connes:Distance} A. Connes,  
``Compact metric spaces, Fredholm modules, and hyperfiniteness,"  
\textit{Ergodic Theory Dynam. Systems} \textbf{9}(2), 207–220 (1989).  
\href{https://doi.org/10.1017/S0143385700004934}{doi:10.1017/S0143385700004934}.


\bibitem{Martinettidistance} P. Martinetti,  
``Distances en géométrie non commutative,"  
PhD thesis, Université de Provence, 2001.
\href{https://arxiv.org/abs/math-ph/0112038}{https://arxiv.org/abs/math-ph/0112038}

\bibitem{Iochum:Krajewski:Martinetti} B. Iochum, T. Krajewski, and P. Martinetti,  
``Distances in finite spaces from noncommutative geometry,"  
\textit{J. Geom. Phys.} \textbf{37}(1-2), 100–125 (2001).
\href{ 	
	https://doi.org/10.48550/arXiv.math-ph/0112038}{ 	
	doi:10.48550/arXiv.math-ph/0112038}.


\bibitem{KKtheory} D. Bailin and A. Love,  
``Kaluza-Klein Theories,"  
\textit{Rep. Prog. Phys.} \textbf{50}(9), 1087–1108 (1987).  
\href{https://doi.org/10.1088/0034-4885/50/9/001}{doi:10.1088/0034-4885/50/9/001}.

		
\bibitem{Chamseddine:2007oz} A. H. Chamseddine, A. Connes, and M. Marcolli,  
``Gravity and the Standard Model with Neutrino Mixing,"  
\textit{Adv. Theor. Math. Phys.} \textbf{11}, 991–1089 (2007).  
\href{https://doi.org/10.4310/ATMP.2007.v11.n6.a3}{doi:10.4310/ATMP.2007.v11.n6.a3}.

	
	
	
\bibitem{Boyle:2020} L. Boyle and S. Farnsworth,  
``The Standard Model, the Pati-Salam Model, and 'Jordan Geometry',"  
\textit{New J. Phys.} \textbf{22}, 073023 (2020).  
\href{https://doi.org/10.1088/1367-2630/ab9709}{doi:10.1088/1367-2630/ab9709}.

	
\bibitem{townsandreview} P. K. Townsend,  
``The Jordan Formulation of Quantum Mechanics: A Review,"  
arXiv:1612.09228 (2016), \href{https://arxiv.org/abs/1612.09228}{https://arxiv.org/abs/1612.09228}.

	
	
	
	
\bibitem{Dungen} K. van den Dungen and W. van Suijlekom,  
``Particle Physics from Almost-Commutative Spacetime,"  
\textit{Rev. Math. Phys.} \textbf{24}(9), 1230004 (2012).  
\href{https://doi.org/10.1142/S0129055X1230004X}{doi:10.1142/S0129055X1230004X}.

		
\bibitem{Farnsworth_2015} L. Boyle and S. Farnsworth,  
``Rethinking Connes' Approach to the Standard Model of Particle Physics via Non-Commutative Geometry,"  
\textit{New J. Phys.} \textbf{17}(2), 023021 (2015).  
\href{https://doi.org/10.1088/1367-2630/17/2/023021}{doi:10.1088/1367-2630/17/2/023021}.

		
		
		
		
		
\bibitem{landi1997} G. Landi,  
``An Introduction to Noncommutative Spaces and Their Geometry,"  
(Lecture Notes in Physics Monographs, Springer, 2003).  
\href{https://doi.org/10.1007/3-540-14949-X}{doi:10.1007/3-540-14949-X}.


\bibitem{Bourbaki} Bourbaki, `` Algèbre: Chapitres 1 à 3'', (Springer Berlin, Heidelberg 2007).  ISBN 978-354-0338-505








		\bibitem{Boyle:2014wba} L.~Boyle and S.~Farnsworth,  
		``Non-Commutative Geometry, Non-Associative Geometry and the Standard Model of Particle Physics,"  
		\textit{New J. Phys.} \textbf{16} (12), 123027 (2014),  
		\href{https://doi.org/10.1088/1367-2630/16/12/123027}{doi: 10.1088/1367-2630/16/12/123027}.
		
		
		

		
\bibitem{toro} L.~A.~Wills-Toro,  
``Classification of Some Graded Not Necessarily Associative Division Algebras I,"  
\textit{Commun. Algebra.} \textbf{42} (12), 5019–5049 (2014),  
\href{https://doi.org/10.1080/00927872.2013.830730}{doi:10.1080/00927872.2013.830730}.

		
\bibitem{Fredy} F.~Giraldo,  
``Extended Higgs Sector on Noncommutative Geometry and Nonassociative Lepton Chiral Symmetry,"  
PhD thesis, Universidad de Antioquia (2021).
\href{https://bibliotecadigital.udea.edu.co/bitstream/10495/29954/1/JimenezFredy_2021_HiggsNoncommutativeGeometry.pdf}{JimenezFredy_2021_HiggsNoncommutativeGeometry}.

	
	
\bibitem{baez} J.~C.~Baez,  
``The Octonions,"  
\textit{Bull. Amer. Math. Soc.} \textbf{39}, 145–205 (2002),  
\href{https://doi.org/10.1090/S0273-0979-01-00934-X}{doi: 10.1090/S0273-0979-01-00934-X}.

\bibitem{gari1} S.~Garibaldi,  
``Structurable Algebras and Groups of Type \(E_6\) and \(E_7\),"  
\textit{J. Algebra.} \textbf{236}(2), 651–691 (2001),  
\href{https://doi.org/10.1006/jabr.2000.8514}{doi: 10.1006/jabr.2000.8514}.

	
\bibitem{boylef4} L.~Boyle,  
``The Standard Model, The Exceptional Jordan Algebra, and Triality," 
arXiv:2006.16265 [hep-th] (2020),  
\href{https://doi.org/10.48550/arXiv.2006.16265}{doi:10.48550/arXiv.2006.16265}.

	
	


	
	\bibitem{Wulkenhaar_1997} R. Wulkenhaar, ``The Standard Model within Non-Associative Geometry," \textit{Phys. Lett. B.} \textbf{390}(1–4), 119–127 (1997), \href{http://dx.doi.org/10.1016/S0370-2693(96)01336-6}{doi:10.1016/S0370-2693(96)01336-6}.
	



	\bibitem{Hassanzadeh_2015} M. Hassanzadeh, I. Shapiro, S. Sütlü, ``Cyclic homology for Hom-associative algebras,''  \textit{J. Geom. Phys.}
{\bf 98}, 40–56 (2015), 
\href{http://dx.doi.org/10.1016/j.geomphys.2015.07.026}{http://dx.doi.org/10.1016/j.geomphys.2015.07.026}.
	


		
\bibitem{Akrami_2004} S. E. Akrami and S. Majid, \emph{Braided Cyclic Cocycles and Nonassociative Geometry}, ``Journal of Mathematical Physics," \textbf{45}(10), 3883–3911 (2004), \href{http://dx.doi.org/10.1063/1.1787621}{doi:10.1063/1.1787621}.


	
\bibitem{carotenuto2019} A.~Carotenuto, L.~D\text{{\`a}}browski, and M.~Dubois-Violette, ``Differential Calculus on Jordan Algebras and Jordan Modules," \textit{Lett. Math. Phys.} \textbf{109}(1), 113–133 (2019), \href{https://doi.org/10.1007/s11005-018-1102-z}{doi:10.1007/s11005-018-1102-z}.

	
	
\bibitem{DV:2016} M.~Dubois-Violette, ``Exceptional Quantum Geometry and Particle Physics," \textit{Nuclear Phys. B.} \textbf{912}, 426–449 (2016), \href{https://doi.org/10.1016/j.nuclphysb.2016.04.018}{doi:10.1016/j.nuclphysb.2016.04.018}.

	
\bibitem{DV:2019} M.~Dubois-Violette and I.~Todorov, ``Exceptional Quantum Geometry and Particle Physics II," \textit{Nuclear Phys. B}, \textbf{938}, 751–761 (2019), \href{https://doi.org/10.1016/j.nuclphysb.2018.12.012}{doi:10.1016/j.nuclphysb.2018.12.012}.

	
	
\bibitem{Farnsworth:2013nza} S.~Farnsworth, L.~Boyle, ``Non-Associative Geometry and the Spectral Action Principle," \textit{JHEP} \textbf{07}, 023 (2015), \href{https://doi.org/10.1007/JHEP07(2015)023}{doi:10.1007/JHEP07(2015)023}.

	
\bibitem{Farnsworth:2014vva} S.~Farnsworth and L.~Boyle, ``Rethinking Connes' approach to the standard model of particle physics via non-commutative geometry," \textit{New J. Phys.} \textbf{17} (2), 023021 (2015), \href{https://doi.org/10.1088/1367-2630/17/2/023021}{doi:10.1088/1367-2630/17/2/023021}.

	
	
\bibitem{Farnsworth:2020ozj} S.~Farnsworth, ``The geometry of physical observables," \textit{J. Math. Phys.} \textbf{61}, 101702 (2020), \href{https://doi.org/10.1063/5.0021707}{doi:10.1063/5.0021707}.

	
	
	
	\bibitem{ShaneThesis} S.~Farnsworth, 		``Standard model physics and beyond
		from non-commutative geometry," PhD thesis, University of Waterloo, (2015).
	\href{https://uwspace.uwaterloo.ca/items/d4ad7fda-8148-4fd0-9d65-ae7a1db648ec}{uwspace.uwaterloo.ca}.
	
\bibitem{Besnard_2022} F. Besnard and S. Farnsworth, ``Particle models from special Jordan backgrounds and spectral triples," Journal of Mathematical Physics, \textbf{63}, 103505 (2022), \href{http://dx.doi.org/10.1063/5.0107136}{doi:10.1063/5.0107136}.

\bibitem{Barnes:2016cjm} G. Barnes, A. Schenkel, and R. Szabo, ``Working with Nonassociative Geometry and Field Theory," \textit{PoS.} \textbf{EMPG-16-02}, 081 (2016), \href{https://doi.org/10.22323/1.263.0081}{doi:10.22323/1.263.0081}.
	
	
\bibitem{Freudenthal} H.~Freudenthal, ``Lie groups in the foundations of geometry," \textit{Adv. Math.}
\textbf{1}(2), 145-190 (1964), \href{https://doi.org/10.1016/0001-8708(65)90038-1}{doi:10.1016/0001-8708(65)90038-1}
	
	\bibitem{Furey_2018} C.~Furey, ``Three generations, two unbroken gauge symmetries, and one eight-dimensional algebra,"
		volume={785},
		ISSN={0370-2693},
		\textit{Phys. Lett. B.} 84-89 (2018), \href{http://dx.doi.org/10.1016/j.physletb.2018.08.032}{doi:10.1016/j.physletb.2018.08.032}.
	
	
	
	
	
\bibitem{Bhatt_2022} V. Bhatt, R. Mondal, V. Vaibhav, and T. P. Singh, ``Majorana neutrinos, exceptional Jordan algebra, and mass ratios for charged fermions," \textit{J. Phys. G. Nuc. Par. Phys.} \textbf{49}(4), 045007 (2022), \href{http://dx.doi.org/10.1088/1361-6471/ac4c91}{doi:10.1088/1361-6471/ac4c91}

\bibitem{Patel_2023} A. A. Patel, T. P. Singh, ``CKM Matrix Parameters from the Exceptional Jordan Algebra," \textit{Universe,} \textbf{9}(10), 440 (2023), \href{http://dx.doi.org/10.3390/universe9100440}{doi:10.3390/universe9100440}.
	
	
	
	\bibitem{jacob1968} N. Jacobson, ``Structure and Representations of Jordan Algebras," (American Mathematical Society, 1968).
ISBN: 978-082-1831-793.
	
		\bibitem{Schafer} R.~Schafer, ``An introduction to nonassociative algebras," (Academic Press, 1966) ISBN-13: 978-048-6688-138
		
		\bibitem{Ramond} P.~Ramond, ``Introduction to Exceptional
		Lie Groups and
		Algebras," \textbf{CALT-68-577}, (1976)
\href{https://lib-extopc.kek.jp/preprints/PDF/1977/7703/7703037.pdf}{KEK Scanned Document}.		
		
	
\bibitem{Chamseddine:2012fk} A.~H. Chamseddine and A.~Connes, ``Resilience of the spectral standard model,'' \textit{JHEP} \textbf{09}, 104 (2012), \href{https://doi.org/10.1007/JHEP09(2012)104}{doi:10.1007/JHEP09(2012)104}.




	
	\bibitem{jordanneumannwigner}	P. Jordan, J. von Neumann and E. P. Wigner, ``On an Algebraic Generalization of the Quantum Mechanical Formalism," \textit{Ann. Math.} \textbf{35}(1), 29-64 (1934). \href{https://doi.org/10.2307/1968117}{doi:10.2307/1968117}.
	

	\bibitem{Eilenberg}S. Eilenberg, ``Extensions of general algebras," \textit{Ann. Soc. Math. Pol.} \textbf{21},
125-134, (1948). \href{https://api.semanticscholar.org/CorpusID:60019734}{https://api.semanticscholar.org/CorpusID:60019734}. 







	\bibitem{Kashuba} 	I. Kashuba, S. Osvienko, I. Shestakov, ``Representation type of Jordan algebras," \textit{Adv. Math.} \textbf{226}, 1-35, (2011). \href{https://doi.org/10.1016/j.aim.2010.07.003}{doi:10.1016/j.aim.2010.07.003}.
	
	
	
	\bibitem{ConnesJ} A. Connes,
	``Noncommutative geometry and Reality," \textit{J. Math. Phys.} \textbf{36}, 6194–6231 (1995),\href{https://doi.org/10.1063/1.531241}{doi:10.1063/1.531241}.
	
	


\bibitem{todorov} I. Todorov and M. Dubois-Violette, ``Deducing the symmetry of the standard model from the automorphism and structure groups of the exceptional Jordan algebra," \textit{Int. J.
Mod. Phys. A} \textbf{33}(20), 1850118 (2018).
\href{https://doi.org/10.1142/S0217751X1850118X}{doi:10.1142/S0217751X1850118X}. 






\bibitem{tod2} M. Dubois-Violette and I. Todorov, “Exceptional quan
tum geometry and particle physics II,” \textit{Nucl. Phys. B,}
\textbf{938}, 751-761 (2019). \href{https://doi.org/10.1016/j.nuclphysb.2018.12.012}{doi:10.1016/j.nuclphysb.2018.12.012}.

\bibitem{Krasnov_2021} K. Krasnov,
``SO(9) characterization of the standard model gauge group,'' \textit{J.  Math. Phys.} \textbf{62}(2), (2021), \href{http://dx.doi.org/10.1063/5.0039941}{doi:10.1063/5.0039941}



	
	\bibitem{Chamseddine_1997} A. Chamseddine, and A.  Connes, ``The Spectral Action Principle,"
	\textit{Comm. Math. Phys.} \textbf{186}(3), 731–750, (1997). \href{https://doi.org/10.1007/s002200050126}{10.1007/s002200050126}.
		
	


	
	
	
	
	
	

\end{thebibliography}
\end{document}